\documentclass[reprint,amssymb,superscriptaddress, amsmath,aps,prb]{revtex4-2}

\usepackage{graphicx}
\usepackage{dcolumn}
\usepackage{epstopdf}
\usepackage{bm}
\usepackage{float}
\usepackage{amsbsy}
\usepackage{array}
\usepackage{amssymb}
\usepackage{setspace}
\usepackage{subcaption}
\usepackage{color}
\usepackage{lineno}
\usepackage{ulem}
\usepackage[colorlinks,linkcolor=red]{hyperref}

\def\Neel{N$\rm\acute{e}$el}
\def\BiSe{Bi$ _{2} $Se$ _{3}$ }

\begin{document}

\title{Magnetic-order-mediated carrier and phonon dynamics in MnBi$_2$Te$_4$}

\author{L. Cheng$^*$}
\thanks{These authors contributed equally to this work}
\affiliation{State Key Laboratory of Electronic Thin Films and Integrated Devices, University of Electronic Science and Technology of China, Chengdu 610054, China}
\author{T. Xiang$^*$}
\thanks{These authors contributed equally to this work}
\affiliation{State Key Laboratory of Electronic Thin Films and Integrated Devices, University of Electronic Science and Technology of China, Chengdu 610054, China}
\author{J. Qi}
\email{jbqi@uestc.edu.cn} 
\affiliation{State Key Laboratory of Electronic Thin Films and Integrated Devices, University of Electronic Science and Technology of China, Chengdu 610054, China}
\date{\today}

\begin{abstract}
We investigate the quasiparticle dynamics in $\mathrm{MnBi_2Te_4}$ single crystal using the ultrafast optical spectroscopy. Our results show that there exist anomalous dynamical optical responses below the antiferromagnetic (AFM) ordering temperature $T_N$. In specific, we reveal that both the initial carrier decay and recombination processes can be modulated via introducing the AFM order in sub-picosecond and picosecond timescales, respectively. We also discover a long relaxation process emerging below $T_N$ with a timescale approaching to the nanosecond regime, and can be attributed to the $T$-dependent spin-lattice interaction. There also emerges an unusual phonon energy renormalization below $T_N$, which is found to arise from its coupling the spin degree via the exchange interaction and magnetic anisotropy. Our findings provide key information for understanding the dynamical properties of non-equilibrium carrier, spin and lattice in $\mathrm{MnBi_2Te_4}$. 
\end{abstract}
	
\maketitle

Magnetic order is one of the essential elements for manipulating the quantum materials due to its capability of mediating the abundant spin-related phenomena~\cite{Beidenkopf2022,Novoselov2019,RevModPhys.82.2731}, as exemplified in strong correlated systems, magnetic two-dimensional materials, as well as the highly focused magnetic topological materials. Their promising applications, e.g., spintronic devices, photonics, and quantum information, are based on the research of interactions between magnetism and related degrees of freedom~\cite{RevModPhys.82.2731,MacDonald2018}. $\mathrm{MnBi_2Te_4}$, a three-dimensional (3D) antiferromagnetic (AFM) topological insulator (TI), is very suitable for studying the interaction between the magnetic order and fermionic carriers in topological non-trivial band. It has a van der Waals-type layered structure with the space group $ R\bar{3}m$~\cite{Otrokov2019Exper,Lee2013}. Previous theoretical calculation and experiments show that its \Neel\ temperature ($T_N$) is around 24~K, below which the AFM order along crystallographic $c$-axis occurs~\cite{Liu2020, Otrokov2019Exper,Zeugner2019,Gong2019}. Compared to the conventional nonmagnetic TIs, the $\mathrm{MnBi_2Te_4}$ is also expected to be ideal for exploring the exotic topological quantum phenomena such as quantum anomalous Hall effect, axion insulator states etc.~\cite{Hasan2010,Liu2020,Li2019,Zhang2019,Gong2019,Otrokov2019Cal,Otrokov2019Exper,Zeugner2019,Yan2019,Mogi2017,Liu2020}, where it is critical to understand how the charge and lattice degrees of freedom interact with the magnetic order below the transition temperature in this system. The spin-lattice interaction below $T_N$ is found to play the key role for the $A_{1g}^1$ phonon mode with terahertz (THz) frequency deviating from the anharmonic model in both the bulk $\mathrm{MnBi_2Te_4}$ \cite{Padmanabhan2022} and few-layer $\mathrm{MnBi_2Te_4}$ flakes \cite{doi:10.1021/acs.nanolett.1c01719}, where, however, there exists discrepancy of the reported $T$-dependent behaviors. Besides, it is still unclear whether the low-energy phonons with gigahertz (GHz) frequency can be coupled to the spin system ~\cite{Bartram2022}. Consequently, several issues remain to be elucidated in this material system, i.e. (1) interaction between the non-equilibrium carriers and the magnetic order, and (2) the spin-lattice coupling in various timescales or energy-scales.   
	
The ultrafast optical pump-probe  spectroscopy (UOPP) has been proved to be a very effective technique in studying the coupling between different degrees of freedom in ultrafast timescale, including the charge, phonon, as well as spin or magnon. In this work, we perform the temperature-dependent UOPP measurements in $\mathrm{MnBi_2Te_4}$ single crystal to study the carrier and lattice dynamics, which are discovered to be strongly coupled with the AFM order below $T_N$. We discuss quantitatively the relevant interactions contributing to the observed dynamics.
	
\begin{figure}\centering
	\includegraphics[width=1\linewidth]{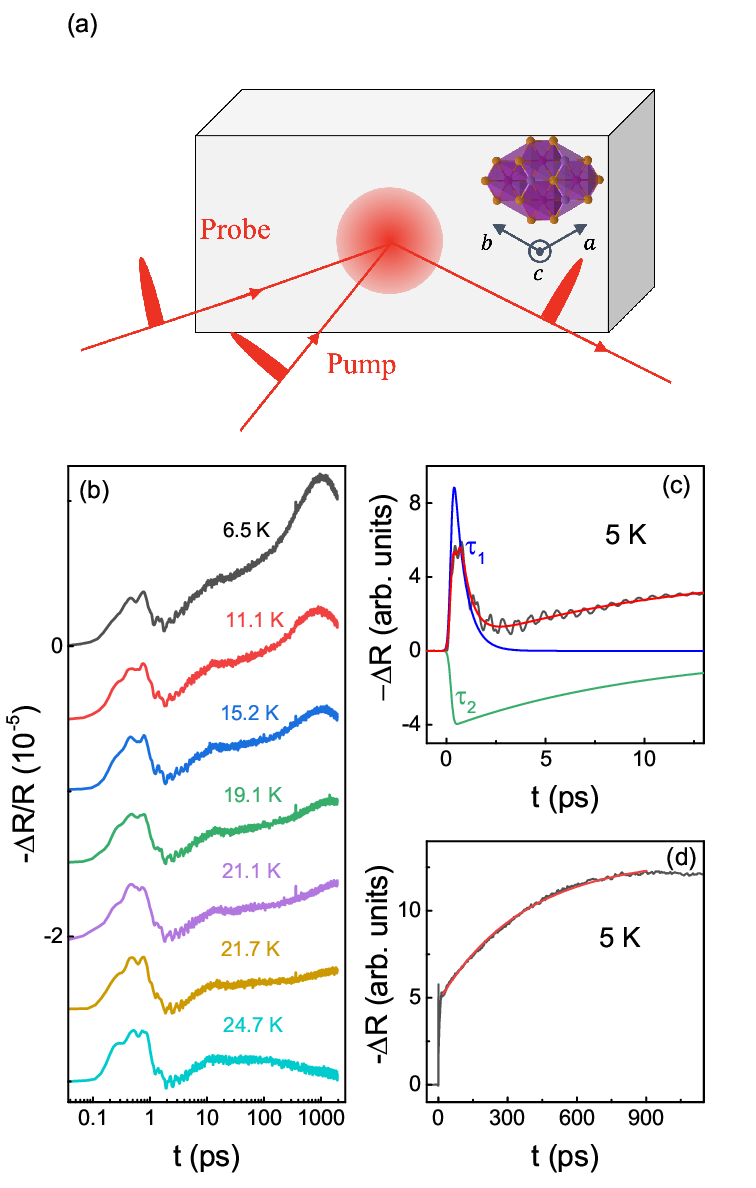}
	\captionsetup{justification=raggedright}
	\caption{(a) Schematic picture of the ultrafast optical pump-probe spectroscopy setup. (b) Transient reflectivity $\Delta R(t)/R$ as a function of temperature for bulk $\mathrm{MnBi_2Te_4}$ single crystal. (c) Fitting of the non-oscillatory signal at 5~K in short timescale. (d) Fitting of the long-rising component at 5~K in long timescale.}
	\label{Fig1}
\end{figure}

The single crystal $\mathrm{MnBi_2Te_4}$ sample was synthesized by flux method. In our UOPP experiments, the transient differential reflectivity $\Delta R(t)/R$ was measured using a Ti:sapphire femtosecond (fs) laser with a center wavelength of 800 nm ($\sim$1.51~eV) and a pulse width of $\sim$35~fs~\cite{Qi2010} (see Fig.~\ref{Fig1}(a)). Details about the experimental setup, sample preparation and its characterization are shown in the Supplemental Material.

Figure~\ref{Fig1}(b) shows the measured signal $\Delta R/R$ as a function of temperature. Upon photoexcitation, we observe an instantaneous rise followed by a non-oscillatory relaxation superimposed with oscillatory signals. The relaxation components exhibit several decays with different lifetimes, indicating multiple relaxation channels for the photoexcited carriers. Surprisingly, after $\sim$10 picosecond (ps) below the \Neel\ temperature, there emerges a very slow rising process extending into nanosecond regime, characterized by $\tau_{\rm long}$.

In fact, the signals $\Delta R/R$ within the initial $\sim$10 ps are quite similar with those measured in other topological insulators, e.g., Bi$_2$Se$_3$~\cite{Qi2010,Lai2014,Cheng2014}, where the oscillatory and exponential decay processes could also be found. This is understandable considering that the $\mathrm{MnBi_2Te_4}$ belongs to the 3D topological insulator \BiSe family with layered structure. In these materials, the short-lifetime decay process are usually related to the carrier relaxations, such as the carrier cooling or recombination, while the oscillatory components represents for the coherent phonon or other collective excitation. They will be discussed in details below.	

First, we focus on the relaxation processes within the initial $\sim$10~ps. To quantify these decays, we fit the signals with following formula~\cite{Wang2015,Hilton2002}:
\begin{equation}
	\dfrac{\Delta R(t)}{R}=\left(\sum_{i=1,2}A_{i}e^{-\frac{t}{\tau_{i}}}+C\right)\otimes G(t),
\end{equation}
where $A_{j}$ and $\tau_{j}$ ($j=1,2$) are the amplitudes and relaxation times, respectively. $C$ is a constant represents for long-lifetime processes, and $G(t)$ is a Gaussian function standing for the pump-probe cross correlation. The fitting quality is very good, as manifested by Fig.~\ref{Fig1}(c), where we can see that $A_1<0$ with $\tau_1$ having a sub-ps timescale (blue curve), and $A_2>0$ with $\tau_2$ having a timescale of few picoseconds (green curve).
\begin{figure}
	\centering
	\includegraphics[width=0.8\linewidth]{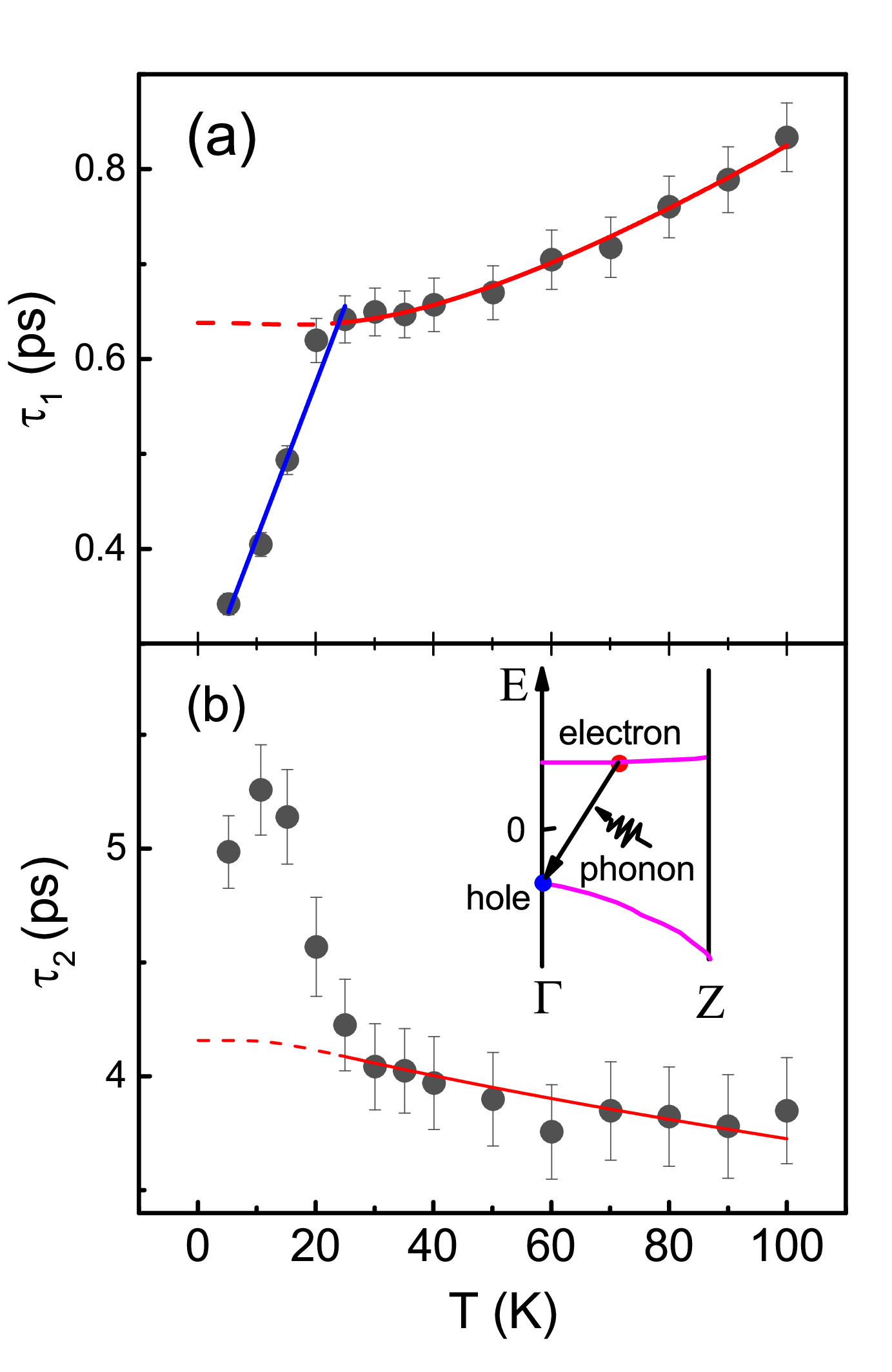}
	\captionsetup{justification=raggedright}
	\caption{The decay times $\tau_{1}$ (a) and $\tau_{2}$ (b) as a function of temperature. The red solid curve in (a) is the TTM fitting at $T>T^*$ and the dashed curve is the extension to below $T^*$. The blue solid curve is the fitting via electron-magnon interaction for $T<T^*$. In (b), the red curve is the fitting based on the phonon-assisted e-h recombination process, which is schematically shown by the inset.}
	\label{Fig2}
\end{figure}

The decay time $\tau_{1} $ as a function of temperature is shown in Fig.~\ref{Fig2}(a). $\tau_{1}$ above $T_N$ with a sub-ps timescale is usually attributed to electron-phonon scattering~\cite{Cheng2014,Dai2015,Cheng2022}. Quantitatively, it could be described by the two-temperature model (TTM)~\cite{Cheng2014,Dai2015,Cheng2022} (see Supplemental Material), which was previously used to study the relaxation dynamics of photoexcited carriers in metals or semimetals.

As seen in Fig.~\ref{Fig2}(a), the fitting quality via TTM is excellent and proves its validity to describe the fast relaxation process characterized by $\tau_1$ at $T>T_N$. We can obtain the electronic specific heat coefficient $\gamma= 4.5$~Jm$^{-3}$K$^{-2}$ and the e-ph coupling constant $g_{\infty}=8.9\times 10^{15} $ Wm$^{-3}$K$ ^{-1}$. These fitted parameters are close to those of the 3D topological insulators in Bi$_2$Se$_3$ family~\cite{Cheng2014}, indicating their similar electron-phonon interaction. 

Particularly, we notice that $\tau_{1}$ decreases rapidly below the {\Neel} temperature. The sudden change of $\tau_{1}$ near $T_N$ clearly cannot be explained by the conventional TTM. Due to the AFM phase transition appearing below $T_N$, the magnetic order should strongly affect the carrier relaxation~\cite{Thielemann-Kuhn2017a,Rettig2012}. Specifically, at $T<T_N$, values of $\tau_{1}$ are clearly smaller than those expected by the conventional TTM, suggesting that formation of the AFM order opens a new relaxation channel for the hot electrons, which might arise from the electron-magnon interaction \cite{PhysRevB.78.174422}. This type of relaxation could be estimated by simply considering the analogy between phonons and magnons. The associated decay time $\tau_{em}$ can be estimated by~\cite{PhysRevB.78.174422}:
\begin{equation}
	\tau_{em}=\dfrac{5\pi k_{B}T_{e}}{3\hbar \lambda_{m}\omega_{m}^{2}}.
\end{equation}
Here $\lambda_{m}$ and $\omega_{m}$ are the electron-magnon coupling constant and magnon cut-off frequency, respectively. We, thus, can fit the values of $\tau_1$ below $T_N$ using the formula $\tau^{-1}=\tau_{em}^{-1}+\tau_c^{-1}$, where $\tau_c^{-1}$ represents the contribution from temperature-insensitive relaxation channel, which is often attributed to the e-ph coupling at low temperatures~\cite{Cheng2014,groeneveld1995}. The fitted results agree very well with the experimental data, as shown in Fig.~\ref{Fig2}(a). We further estimate $\lambda_m\omega_m^2$ to be $42.1 $~THz$^{2}$. Although there is no known parameter value(s) for comparison due to the lack of related work from other research groups, the obvious AFM-related $T$-dependence of $\tau_{1}$ demonstrates that the electron-magnon interaction could be vital.

$\tau_{2}$ as a function of temperature, characterizing the decay process with a timescale of several ps, is shown in Fig.~\ref{Fig2}(b). As is known, after the e-ph thermalization, the nonequilibrium electrons (holes) can cool down and accumulate at the minimum (maximum) of conduction (valence) band. The subsequent relaxation involves the recombination processes. Since the direct recombination normally takes a long time with nanosecond timescale \cite{Schroder2005}, it cannot account for the $T$-dependent $\tau_2$. However, the phonon-assisted electron-hole (e-h) recombination \cite{Sheu2013,Dai2015,Cheng2022}, where the electron and hole recombine with the assistance of e-ph scattering between the electron and hole pockets, has exactly the same timescale as that of $\tau_2$. Actually, in the electronic structure of single crystal $\mathrm{MnBi_2Te_4}$, the conduction band can be very flat along some specific momentum direction within the Brillouin zone, i.e. bands along $\Gamma-Z$ in the inset of Fig.~\ref{Fig2}(b) \cite{Li2021}, which is in favor of such indirect recombination. Therefore, we can use the phonon-assisted e-ph recombination model to describe the decay time $\tau_{2}$ as
\begin{equation}
	\frac{1}{\tau_{2}}=A\frac{x}{\sinh^{2}x}+\dfrac{1}{\tau_{0}},
\end{equation}
where $x=\hbar\omega_r/2k_{B}T$, and $\omega_r$ corresponds to the average frequency of phonon involved in the e-h recombination. $\tau_{0}$ represents a temperature independent recombination time which relies on the impurities or defects of the sample. $A$ is a parameter related to the density of states in the electronic energy bands and the matrix elements for interband e-h scattering. It can be seen from Fig.~2(b) that the fitted results agree quite well with the experimental $\tau_{2}$ above the {\Neel} temperature. We further obtain that $\omega/2\pi = 1.38$~THz and $\tau_{0} = 4.16 $~ps. Here, $\omega$ is quite close to frequency of the dominant oscillation observed (as shown in Fig.~\ref{Fig4}(a-c)), i.e. $A^1_{1g}$ mode.
	
Similar to the temperature-dependent $\tau_{1}$, the values of $\tau_{2}$ at $T<T_N$ deviate from the trend predicted by the indirect recombination scenario. Due to existence of the long-range AFM order in this temperature regime, two possible mechanisms might contribute to such temperature-dependence. First, we can consider the interaction between phonon and magnon \cite{Padmanabhan2022}, which can renormalize the phonon frequency and change its population, and hence is able to affect the phonon-assisted recombination. Second, the magnon-assisted recombination can also contribute to the carrier relaxation, where the magnon plays similar role as the phonon, i.e. compensates the conservation of energy and momentum in the recombination process. However, in the second case magnon should open a new relaxation channel that will accelerate the recombination process. This is clearly not consistent with our observation. Therefore, we believe that the phonon-magnon interaction, rather than the magnon-assisted recombination, dominates the relaxation characterized by $\tau_{2}$ at $T<T_N$. Renormalization of the phonon energy is indeed observed in our experiment, as discussed in detail below. 

\begin{figure}
	\centering
	\includegraphics[width=1\linewidth]{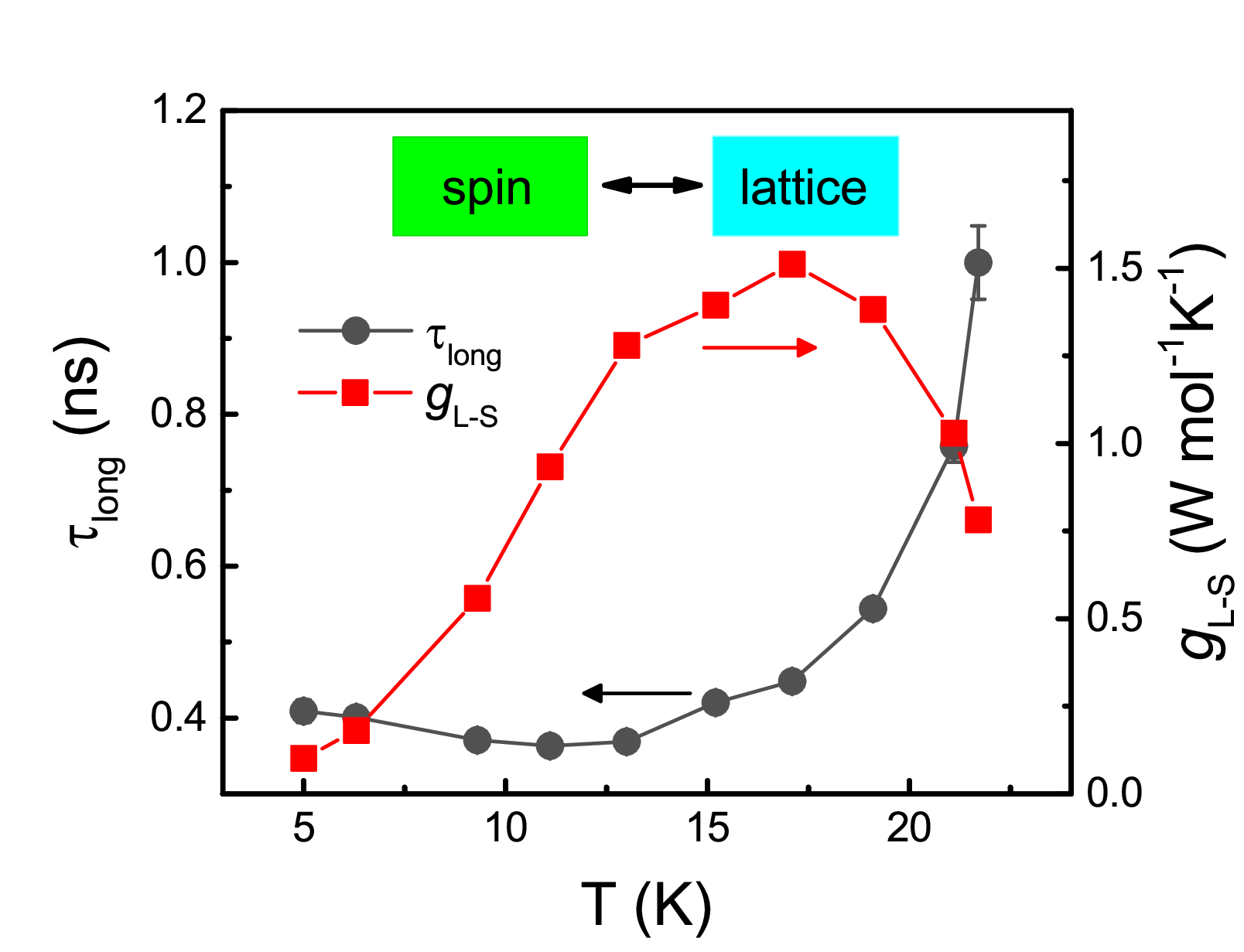}
	\captionsetup{justification=raggedright}
	\caption{$\tau_{long}$ and $g_{L-S}$ as functions of temperature.}
	\label{Fig3}
\end{figure}

\begin{figure}
	\centering
	\includegraphics[width=1\linewidth]{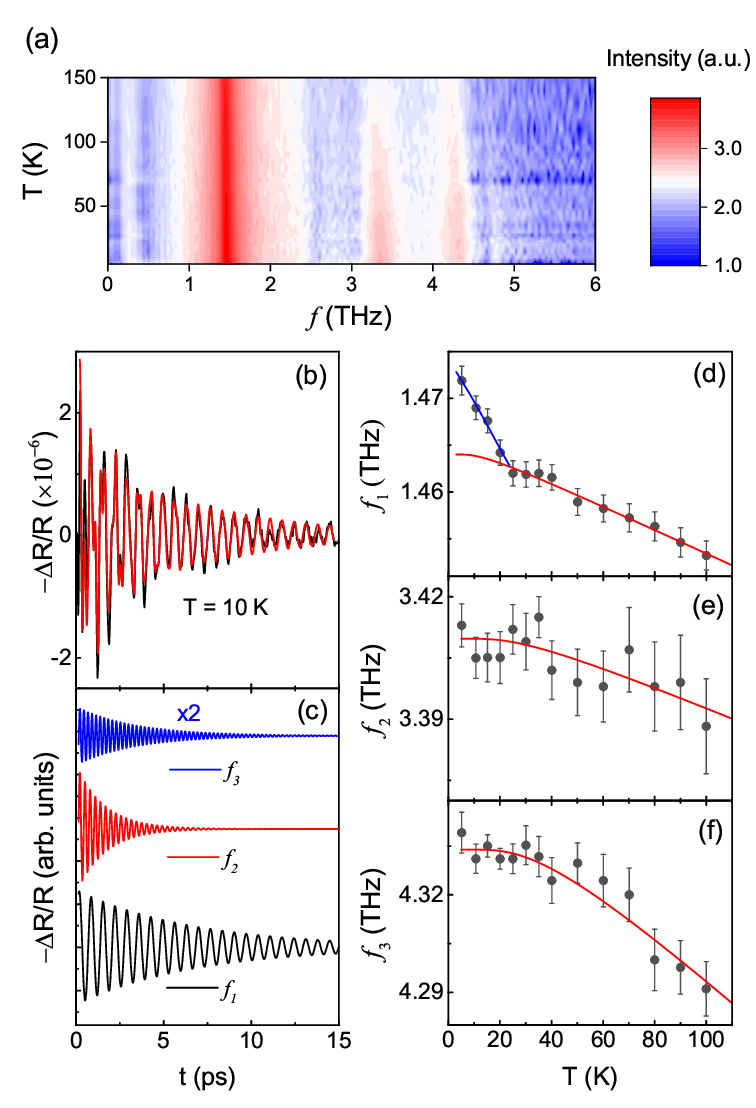}
	\captionsetup{justification=raggedright}
	\caption{(a) Fourier transform spectra of the extracted oscillations as a function of temperature (map color rescaled via log$_2$). (b) Extracted oscillation at 10 K (black) and the fitted curve (red) using the damped harmonic oscillators. (c) The three oscillation components with different frequencies derived from the fitting. (d-f) The frequencies of three phonon modes obtained via the fitting as a function of temperature. The red lines are fitted curves using Eq.~(\ref{eqh}). The blue line in (d) is the fit via Eq.~(\ref{alpha}).}
	\label{Fig4}
\end{figure}

As mentioned earlier, after $\sim$10 ps there emerges a long-rising process below \Neel\ temperature characterized by $\tau_{\rm long}$, which should originate from the AFM order~\cite{Qi2012,Vaterlaus1991,Yuan2019,Chia2007}. Its $T$-dependent timescale spans from hundreds of ps to nanosecond, and is consistent with previous observation associated with the spin-lattice interaction in several other magnetic materials~\cite{Qi2012,Vaterlaus1991,Chia2007}. This type of interaction could be phenomenologically described by the energy exchange between spin and lattice subsystems~\cite{Qi2012}. Here, $T^{\rm L} $ and $T^{\rm S}$ representing the temperatures characterizing lattice and spin reservoirs, respectively. Their coupling can be effectively described by a parameter $g_{\rm L-S}$, which indicates the coupling strength or energy transfer rate. Then the relaxation time of non-equilibrium coupled spin-lattice system can be characterized by a decay time \cite{Qi2012}:
	\begin{equation}
		\label{taulong}
		\tau_{\rm long}^{-1}\propto g_{\rm L-S}(T)\left[\dfrac{1}{C^{\rm L}(T)}+\dfrac{1}{C^{\rm S}(T)}\right].
	\end{equation}
Here, the $ C^{\rm L,S} $ are the phonon and spin heat capacity of the two reservoirs. Values of $\tau_{\rm long}$ can be obtained via the single exponential fit to the experimental data with long timescale, as shown in Fig.~\ref{Fig3}, where a divergence-like behavior near $T_N$ can be seen. Since $C^S$ peaks around $T_N$, such phenomenon can be understood via Eq.(\ref{taulong}). 

According to the $\tau_{\rm long}$, we can further obtain the coupling coefficient $g_{\rm L-S}(T)$ for $\mathrm{MnBi_2Te_4}$ below $T_N$, using the specific heat $ C^{\rm L}$ and $ C^{\rm S}$ extracted from previous heat capacity measurements~\cite{Ning2020,Zeugner2019}. Surprisingly, $g_{\rm L-S}$ does not show a monotonic behavior as a function of temperature below $T_N$, and it maximizes around a critical temperature $T^*\sim17$~K. Phenomenologically, below $T^*$ its increasing with $T$ could be attributed to the increment of phonon population, while above $T^*$ its reduction with increasing $T$ should arise from the magnetization $M(T)$ vanishing gradually when approaching to $T_N$. Moreover, due to the long-lifetime of this spin-lattice interaction process, the involved collective excitation in the scatterings should have low-energy scales, e.g. acoustic phonon and sub-meV magnon. Since the AFM order is along the out-of-plane direction, its associated magnetic excitation should have strongest coupling with modes vibrating along the $c$-axis.  

Fourier transform spectra of the oscillations extracted from the non-oscillatory $\Delta R/R$ components is shown in Fig.~\ref{Fig4}(a). Three coherent optical phonon modes in $\mathrm{MnBi_2Te_4}$ can be clearly seen, and manifested by a pronounced peak, $f_1\sim1.5$~THz, and two smaller peaks, $f_2\sim 3.4$~THz and $f_3\sim 4.3$~THz, respectively. These modes are consistent with the previous Raman spectroscopy studies~\cite{Padmanabhan2022,doi:10.1021/acs.nanolett.1c01719,Aliev2019}, and can be assigned as $A_{1g}^{1}$ ($f_1$), $A_{1g}^{2}$ ($f_2$) and $A_{1g}^{3}$ ($f_3$) modes, respectively.

The data could be fitted well by three damped harmonic oscillators~\cite{LiuYP2020}. A typical example is demonstrated in Fig.~\ref{Fig4}(b-c). The obtained temperature-dependent frequencies $f_j(T)$~($j=1,2,3$) are illustrated in Fig.~\ref{Fig4}(d-f), which show clear softening with increasing temperature and usually can be explained by the anharmonic phonon model ~\cite{PhysRevB.28.1928,PhysRevB.29.2051,LiuYP2020}, where the $T$-dependent phonon frequency $\omega_A(=2\pi f_A)$ is given by:
\begin{equation}
	\label{eqh}
	\omega_A\left(T\right)=\omega_0+a\left[1+2n_B\left(\frac{\omega_0}{2},T\right)\right],
\end{equation}
where $\omega_0$ is the intrinsic frequency, and $a$ is the fitting parameter. $n_B(\omega,T)=[e^{\hbar\omega/k_BT}-1]^{-1}$ is the Bose-Einstein distribution function. We can see that the $T$-dependent phonon frequencies follow this model in general. However, as shown in Fig.~\ref{Fig4}(d), it can only describe the $T$-dependent $f_1$ for temperatures larger than $\sim T_N$, while it fails to predict the behavior of $f_1(T)$ at lower temperatures. Such result agrees with the measurements via Raman spectroscopy~\cite{Padmanabhan2022,doi:10.1021/acs.nanolett.1c01719}. Since $A_{1g}^{1}$ mode represents the interlayer atomic vibrations along the crystallographic $c$-axis, modulation of the interlayer exchange interaction and anisotropy constant becomes easily achievable. This leads to the spin-lattice coupling effect, which will renormalize the energy of related phonon or magnon. 

We can understand such phonon renormalization by incorporating the linear-chain model in MnBi$_2$Te$_4$~\cite{PhysRevX.11.011003}, i.e. the Hamiltonian with a phonon can be written as:
\begin{equation}
	H=\frac{1}{2}\omega_0^2u^2+J\sum_{i=1}^{N-1}\vec{M}_i\cdot\vec{M}_{i+1}-\frac{1}{2}K\sum_{i=1}^N(\vec{M}_i\cdot\hat{z})^2,
\end{equation} 
where $u$ is the atomic displacement. $J$ is the exchange interaction. $\vec{M}_i$ is the magnetization in the ith layer. $K~(>0)$ is easy-axis anisotropy energy along $z$ direction (or $c$-axis). $N$ is the number of septuple layer. We then can solve the phonon frequency for $f_1$ mode via $\omega_1^2=d^2H/du^2$:
\begin{equation}
	\omega_1^2(T)=\omega_0^2+\frac{d^2J}{du^2}\sum_{i=1}^{N-1}\vec{M}_i\cdot\vec{M}_{i+1}-\frac{1}{2}\frac{d^2K}{du^2}\sum_{i=1}^N(\vec{M}_i\cdot\hat{z})^2.
\end{equation}
Under the approximation of small displacement, $J$ and $K$ can be written as $J(u)=J_0+J'u+\frac{1}{2}J''u^2$ and $K(u)=K_0+K'u+\frac{1}{2}K''u^2$. Here, the linear terms in $u$ do not affect the phonon frequency. Therefore, we can expect the $T$-dependence of $\omega_1$ in the low-temperature regime to be:
\begin{equation}
	\omega_1^2=\omega_A^2+J'' \sum_{i=1}^{N-1}\vec{M}_i\cdot\vec{M}_{i+1}-\frac{1}{2}K''\sum_{i=1}^N(\vec{M}_i\cdot\hat{z})^2.
	\label{omega_1}
\end{equation}

In order to catch the main physics, we only consider a simplest case \cite{npjQuantumMaterials5902020}, where $|\vec{M}_i(T)|=M(T)$, and $M_i$ is parallel or antiparallel with the out-of-plane $\hat{z}$ direction (easy axis). Then, Eq.~\ref{omega_1} can be rewritten as: 
\begin{equation}
	\omega_1^2=\omega_A^2-\alpha M^2(T),
	\label{alpha}
\end{equation}
where $\alpha$ is given by: $\alpha=(N-1)J''-\frac{1}{2}NK''$. If $M(T)$ takes a form: $M(T)=M_s(1-T/T_N)^\beta$ \cite{PhysRevX.11.011003}, we can easily fit $f_1(T)$ using the Eq.~(14). As seen from Fig.~4(c), an excellent fit can be obtained with $\beta=0.53\pm0.08$. Such $\beta$ value is consistent very well with the results measured by the reflectance magnetic circular dichroism \cite{PhysRevX.11.011003}. Clearly, Eq.~\ref{alpha} indicates that the competition between $J''$ and $K''$ terms can lead to either $\alpha>0$ or $\alpha<0$. Due to $J$ and $K$ having thickness dependence, we can understand the contradicting softening and hardening effect observed in the few-layer and bulk samples ~\cite{Padmanabhan2022,doi:10.1021/acs.nanolett.1c01719}, respectively. The spin-lattice coupling dominating the phonon renormalization may also inhibit the anharmonic effect, as manifested by the decay of coherent $A_{1g}^1$ mode (see the Supplemental Material). 

We note that the anomalous $T$-dependent $f_1$ below $T_N$ could also be deduced by the phonon-assisted recombination characterized by $\tau_2$, where the $A_{1g}^{1}$ phonon mode participates as the indirect scattering media and $\tau_2$ shows anomaly near $T_N$.  Unfortunately, due to the low signal-to-noise ratio, it is hard to discern if there is any anomaly appearing near $T_N$ in the $T$-dependent phonon modes $f_2$ and $f_3$, as seen in Fig.~\ref{Fig4}(d-e).
	 
In conclusion, we performed ultrafast transient reflectivity measurements on $\mathrm{MnBi_2Te_4}$. The experimental results show that the ultrafast non-equilibrium carrier and phonon dynamics can be strongly coupled to the long-range AFM order via the electron-magnon, electron-phonon and spin-lattice interactions. In specific, The AFM transition at $T_N$ in MnBi$_2$Te$_4$ creates anomaly in the carrier and coherent phonon relaxation processes. It also causes an unusual phonon hardening of $f_1$ mode relative to the values expected by the anharmonic phonon effect below $T_N$, and demonstrates the strong coupling between the AFM order and phonon.

This work was supported by the National Key R\&D Program of China (Grant No. 2022YFA1403002), the National Natural Science Foundation of China (Grants Nos. 11974070, 11734006, 11925408, 11921004, 12004067, 62027807).
      	

\begin{thebibliography}{43}%
\makeatletter
\providecommand \@ifxundefined [1]{%
 \@ifx{#1\undefined}
}%
\providecommand \@ifnum [1]{%
 \ifnum #1\expandafter \@firstoftwo
 \else \expandafter \@secondoftwo
 \fi
}%
\providecommand \@ifx [1]{%
 \ifx #1\expandafter \@firstoftwo
 \else \expandafter \@secondoftwo
 \fi
}%
\providecommand \natexlab [1]{#1}%
\providecommand \enquote  [1]{``#1''}%
\providecommand \bibnamefont  [1]{#1}%
\providecommand \bibfnamefont [1]{#1}%
\providecommand \citenamefont [1]{#1}%
\providecommand \href@noop [0]{\@secondoftwo}%
\providecommand \href [0]{\begingroup \@sanitize@url \@href}%
\providecommand \@href[1]{\@@startlink{#1}\@@href}%
\providecommand \@@href[1]{\endgroup#1\@@endlink}%
\providecommand \@sanitize@url [0]{\catcode `\\12\catcode `\$12\catcode
  `\&12\catcode `\#12\catcode `\^12\catcode `\_12\catcode `\%12\relax}%
\providecommand \@@startlink[1]{}%
\providecommand \@@endlink[0]{}%
\providecommand \url  [0]{\begingroup\@sanitize@url \@url }%
\providecommand \@url [1]{\endgroup\@href {#1}{\urlprefix }}%
\providecommand \urlprefix  [0]{URL }%
\providecommand \Eprint [0]{\href }%
\providecommand \doibase [0]{https://doi.org/}%
\providecommand \selectlanguage [0]{\@gobble}%
\providecommand \bibinfo  [0]{\@secondoftwo}%
\providecommand \bibfield  [0]{\@secondoftwo}%
\providecommand \translation [1]{[#1]}%
\providecommand \BibitemOpen [0]{}%
\providecommand \bibitemStop [0]{}%
\providecommand \bibitemNoStop [0]{.\EOS\space}%
\providecommand \EOS [0]{\spacefactor3000\relax}%
\providecommand \BibitemShut  [1]{\csname bibitem#1\endcsname}%
\let\auto@bib@innerbib\@empty
\bibitem [{\citenamefont {Bernevig}\ \emph {et~al.}(2022)\citenamefont
  {Bernevig}, \citenamefont {Felser},\ and\ \citenamefont
  {Beidenkopf}}]{Beidenkopf2022}%
  \BibitemOpen
  \bibfield  {author} {\bibinfo {author} {\bibfnamefont {B.~A.}\ \bibnamefont
  {Bernevig}}, \bibinfo {author} {\bibfnamefont {C.}~\bibnamefont {Felser}},\
  and\ \bibinfo {author} {\bibfnamefont {H.}~\bibnamefont {Beidenkopf}},\
  }\href {https://doi.org/10.1038/s41586-021-04105-x} {\bibfield  {journal}
  {\bibinfo  {journal} {Nature}\ }\textbf {\bibinfo {volume} {603}},\ \bibinfo
  {pages} {41} (\bibinfo {year} {2022})}\BibitemShut {NoStop}%
\bibitem [{\citenamefont {Gibertini}\ \emph {et~al.}(2019)\citenamefont
  {Gibertini}, \citenamefont {Koperski}, \citenamefont {Morpurgo},\ and\
  \citenamefont {Novoselov}}]{Novoselov2019}%
  \BibitemOpen
  \bibfield  {author} {\bibinfo {author} {\bibfnamefont {M.}~\bibnamefont
  {Gibertini}}, \bibinfo {author} {\bibfnamefont {M.}~\bibnamefont {Koperski}},
  \bibinfo {author} {\bibfnamefont {A.~F.}\ \bibnamefont {Morpurgo}},\ and\
  \bibinfo {author} {\bibfnamefont {K.~S.}\ \bibnamefont {Novoselov}},\ }\href
  {https://link.aps.org/doi/10.1103/PhysRevB.51.11433} {\bibfield  {journal}
  {\bibinfo  {journal} {Nat. Nanotech.}\ }\textbf {\bibinfo {volume} {14}},\
  \bibinfo {pages} {408} (\bibinfo {year} {2019})}\BibitemShut {NoStop}%
\bibitem [{\citenamefont {Kirilyuk}\ \emph {et~al.}(2010)\citenamefont
  {Kirilyuk}, \citenamefont {Kimel},\ and\ \citenamefont
  {Rasing}}]{RevModPhys.82.2731}%
  \BibitemOpen
  \bibfield  {author} {\bibinfo {author} {\bibfnamefont {A.}~\bibnamefont
  {Kirilyuk}}, \bibinfo {author} {\bibfnamefont {A.~V.}\ \bibnamefont
  {Kimel}},\ and\ \bibinfo {author} {\bibfnamefont {T.}~\bibnamefont
  {Rasing}},\ }\href {https://doi.org/10.1103/RevModPhys.82.2731} {\bibfield
  {journal} {\bibinfo  {journal} {Rev. Mod. Phys.}\ }\textbf {\bibinfo {volume}
  {82}},\ \bibinfo {pages} {2731} (\bibinfo {year} {2010})}\BibitemShut
  {NoStop}%
\bibitem [{\citenamefont {Šmejkal}\ \emph {et~al.}(2018)\citenamefont
  {Šmejkal}, \citenamefont {Mokrousov}, \citenamefont {Yan},\ and\
  \citenamefont {MacDonald}}]{MacDonald2018}%
  \BibitemOpen
  \bibfield  {author} {\bibinfo {author} {\bibfnamefont {L.}~\bibnamefont
  {Šmejkal}}, \bibinfo {author} {\bibfnamefont {Y.}~\bibnamefont {Mokrousov}},
  \bibinfo {author} {\bibfnamefont {B.}~\bibnamefont {Yan}},\ and\ \bibinfo
  {author} {\bibfnamefont {A.~H.}\ \bibnamefont {MacDonald}},\ }\href
  {https://doi.org/10.1038/s41567-018-0064-5} {\bibfield  {journal} {\bibinfo
  {journal} {Nat. Phys.}\ }\textbf {\bibinfo {volume} {14}},\ \bibinfo {pages}
  {242} (\bibinfo {year} {2018})}\BibitemShut {NoStop}%
\bibitem [{\citenamefont {Otrokov}\ \emph
  {et~al.}(2019{\natexlab{a}})\citenamefont {Otrokov}, \citenamefont
  {Klimovskikh}, \citenamefont {Bentmann}, \citenamefont {Estyunin},
  \citenamefont {Zeugner}, \citenamefont {Aliev}, \citenamefont {Ga{\ss}},
  \citenamefont {Wolter}, \citenamefont {Koroleva}, \citenamefont {Shikin},
  \citenamefont {Blanco-Rey}, \citenamefont {Hoffmann}, \citenamefont
  {Rusinov}, \citenamefont {Vyazovskaya}, \citenamefont {Eremeev},
  \citenamefont {Koroteev}, \citenamefont {Kuznetsov}, \citenamefont {Freyse},
  \citenamefont {S{\'{a}}nchez-Barriga}, \citenamefont {Amiraslanov},
  \citenamefont {Babanly}, \citenamefont {Mamedov}, \citenamefont {Abdullayev},
  \citenamefont {Zverev}, \citenamefont {Alfonsov}, \citenamefont {Kataev},
  \citenamefont {B{\"{u}}chner}, \citenamefont {Schwier}, \citenamefont
  {Kumar}, \citenamefont {Kimura}, \citenamefont {Petaccia}, \citenamefont {{Di
  Santo}}, \citenamefont {Vidal}, \citenamefont {Schatz}, \citenamefont
  {Ki{\ss}ner}, \citenamefont {{\"{U}}nzelmann}, \citenamefont {Min},
  \citenamefont {Moser}, \citenamefont {Peixoto}, \citenamefont {Reinert},
  \citenamefont {Ernst}, \citenamefont {Echenique}, \citenamefont {Isaeva},\
  and\ \citenamefont {Chulkov}}]{Otrokov2019Exper}%
  \BibitemOpen
  \bibfield  {author} {\bibinfo {author} {\bibfnamefont {M.~M.}\ \bibnamefont
  {Otrokov}}, \bibinfo {author} {\bibfnamefont {I.~I.}\ \bibnamefont
  {Klimovskikh}}, \bibinfo {author} {\bibfnamefont {H.}~\bibnamefont
  {Bentmann}}, \bibinfo {author} {\bibfnamefont {D.}~\bibnamefont {Estyunin}},
  \bibinfo {author} {\bibfnamefont {A.}~\bibnamefont {Zeugner}}, \bibinfo
  {author} {\bibfnamefont {Z.~S.}\ \bibnamefont {Aliev}}, \bibinfo {author}
  {\bibfnamefont {S.}~\bibnamefont {Ga{\ss}}}, \bibinfo {author} {\bibfnamefont
  {A.~U.~B.}\ \bibnamefont {Wolter}}, \bibinfo {author} {\bibfnamefont {A.~V.}\
  \bibnamefont {Koroleva}}, \bibinfo {author} {\bibfnamefont {A.~M.}\
  \bibnamefont {Shikin}}, \bibinfo {author} {\bibfnamefont {M.}~\bibnamefont
  {Blanco-Rey}}, \bibinfo {author} {\bibfnamefont {M.}~\bibnamefont
  {Hoffmann}}, \bibinfo {author} {\bibfnamefont {I.~P.}\ \bibnamefont
  {Rusinov}}, \bibinfo {author} {\bibfnamefont {A.~Y.}\ \bibnamefont
  {Vyazovskaya}}, \bibinfo {author} {\bibfnamefont {S.~V.}\ \bibnamefont
  {Eremeev}}, \bibinfo {author} {\bibfnamefont {Y.~M.}\ \bibnamefont
  {Koroteev}}, \bibinfo {author} {\bibfnamefont {V.~M.}\ \bibnamefont
  {Kuznetsov}}, \bibinfo {author} {\bibfnamefont {F.}~\bibnamefont {Freyse}},
  \bibinfo {author} {\bibfnamefont {J.}~\bibnamefont {S{\'{a}}nchez-Barriga}},
  \bibinfo {author} {\bibfnamefont {I.~R.}\ \bibnamefont {Amiraslanov}},
  \bibinfo {author} {\bibfnamefont {M.~B.}\ \bibnamefont {Babanly}}, \bibinfo
  {author} {\bibfnamefont {N.~T.}\ \bibnamefont {Mamedov}}, \bibinfo {author}
  {\bibfnamefont {N.~A.}\ \bibnamefont {Abdullayev}}, \bibinfo {author}
  {\bibfnamefont {V.~N.}\ \bibnamefont {Zverev}}, \bibinfo {author}
  {\bibfnamefont {A.}~\bibnamefont {Alfonsov}}, \bibinfo {author}
  {\bibfnamefont {V.}~\bibnamefont {Kataev}}, \bibinfo {author} {\bibfnamefont
  {B.}~\bibnamefont {B{\"{u}}chner}}, \bibinfo {author} {\bibfnamefont {E.~F.}\
  \bibnamefont {Schwier}}, \bibinfo {author} {\bibfnamefont {S.}~\bibnamefont
  {Kumar}}, \bibinfo {author} {\bibfnamefont {A.}~\bibnamefont {Kimura}},
  \bibinfo {author} {\bibfnamefont {L.}~\bibnamefont {Petaccia}}, \bibinfo
  {author} {\bibfnamefont {G.}~\bibnamefont {{Di Santo}}}, \bibinfo {author}
  {\bibfnamefont {R.~C.}\ \bibnamefont {Vidal}}, \bibinfo {author}
  {\bibfnamefont {S.}~\bibnamefont {Schatz}}, \bibinfo {author} {\bibfnamefont
  {K.}~\bibnamefont {Ki{\ss}ner}}, \bibinfo {author} {\bibfnamefont
  {M.}~\bibnamefont {{\"{U}}nzelmann}}, \bibinfo {author} {\bibfnamefont
  {C.~H.}\ \bibnamefont {Min}}, \bibinfo {author} {\bibfnamefont
  {S.}~\bibnamefont {Moser}}, \bibinfo {author} {\bibfnamefont {T.~R.~F.}\
  \bibnamefont {Peixoto}}, \bibinfo {author} {\bibfnamefont {F.}~\bibnamefont
  {Reinert}}, \bibinfo {author} {\bibfnamefont {A.}~\bibnamefont {Ernst}},
  \bibinfo {author} {\bibfnamefont {P.~M.}\ \bibnamefont {Echenique}}, \bibinfo
  {author} {\bibfnamefont {A.}~\bibnamefont {Isaeva}},\ and\ \bibinfo {author}
  {\bibfnamefont {E.~V.}\ \bibnamefont {Chulkov}},\ }\href
  {https://doi.org/10.1038/s41586-019-1840-9} {\bibfield  {journal} {\bibinfo
  {journal} {Nature}\ }\textbf {\bibinfo {volume} {576}},\ \bibinfo {pages}
  {416} (\bibinfo {year} {2019}{\natexlab{a}})}\BibitemShut {NoStop}%
\bibitem [{\citenamefont {Lee}\ \emph {et~al.}(2013)\citenamefont {Lee},
  \citenamefont {Kim}, \citenamefont {Park}, \citenamefont {Chung},
  \citenamefont {Lim}, \citenamefont {Seo},\ and\ \citenamefont
  {Park}}]{Lee2013}%
  \BibitemOpen
  \bibfield  {author} {\bibinfo {author} {\bibfnamefont {D.~S.}\ \bibnamefont
  {Lee}}, \bibinfo {author} {\bibfnamefont {T.~H.}\ \bibnamefont {Kim}},
  \bibinfo {author} {\bibfnamefont {C.~H.}\ \bibnamefont {Park}}, \bibinfo
  {author} {\bibfnamefont {C.~Y.}\ \bibnamefont {Chung}}, \bibinfo {author}
  {\bibfnamefont {Y.~S.}\ \bibnamefont {Lim}}, \bibinfo {author} {\bibfnamefont
  {W.~S.}\ \bibnamefont {Seo}},\ and\ \bibinfo {author} {\bibfnamefont {H.~H.}\
  \bibnamefont {Park}},\ }\href {https://doi.org/10.1039/c3ce40643a} {\bibfield
   {journal} {\bibinfo  {journal} {CrystEngComm}\ }\textbf {\bibinfo {volume}
  {15}},\ \bibinfo {pages} {5532} (\bibinfo {year} {2013})}\BibitemShut
  {NoStop}%
\bibitem [{\citenamefont {Liu}\ \emph {et~al.}(2020{\natexlab{a}})\citenamefont
  {Liu}, \citenamefont {Wang}, \citenamefont {Li}, \citenamefont {Wu},
  \citenamefont {Li}, \citenamefont {Li}, \citenamefont {He}, \citenamefont
  {Xu}, \citenamefont {Zhang},\ and\ \citenamefont {Wang}}]{Liu2020}%
  \BibitemOpen
  \bibfield  {author} {\bibinfo {author} {\bibfnamefont {C.}~\bibnamefont
  {Liu}}, \bibinfo {author} {\bibfnamefont {Y.}~\bibnamefont {Wang}}, \bibinfo
  {author} {\bibfnamefont {H.}~\bibnamefont {Li}}, \bibinfo {author}
  {\bibfnamefont {Y.}~\bibnamefont {Wu}}, \bibinfo {author} {\bibfnamefont
  {Y.}~\bibnamefont {Li}}, \bibinfo {author} {\bibfnamefont {J.}~\bibnamefont
  {Li}}, \bibinfo {author} {\bibfnamefont {K.}~\bibnamefont {He}}, \bibinfo
  {author} {\bibfnamefont {Y.}~\bibnamefont {Xu}}, \bibinfo {author}
  {\bibfnamefont {J.}~\bibnamefont {Zhang}},\ and\ \bibinfo {author}
  {\bibfnamefont {Y.}~\bibnamefont {Wang}},\ }\href
  {https://doi.org/10.1038/s41563-019-0573-3} {\bibfield  {journal} {\bibinfo
  {journal} {Nat. Mater.}\ }\textbf {\bibinfo {volume} {19}},\ \bibinfo {pages}
  {522} (\bibinfo {year} {2020}{\natexlab{a}})}\BibitemShut {NoStop}%
\bibitem [{\citenamefont {Zeugner}\ \emph {et~al.}(2019)\citenamefont
  {Zeugner}, \citenamefont {Nietschke}, \citenamefont {Wolter}, \citenamefont
  {Ga{\ss}}, \citenamefont {Vidal}, \citenamefont {Peixoto}, \citenamefont
  {Pohl}, \citenamefont {Damm}, \citenamefont {Lubk}, \citenamefont {Hentrich},
  \citenamefont {Moser}, \citenamefont {Fornari}, \citenamefont {Min},
  \citenamefont {Schatz}, \citenamefont {Ki{\ss}ner}, \citenamefont
  {{\"{U}}nzelmann}, \citenamefont {Kaiser}, \citenamefont {Scaravaggi},
  \citenamefont {Rellinghaus}, \citenamefont {Nielsch}, \citenamefont {Hess},
  \citenamefont {B{\"{u}}chner}, \citenamefont {Reinert}, \citenamefont
  {Bentmann}, \citenamefont {Oeckler}, \citenamefont {Doert}, \citenamefont
  {Ruck},\ and\ \citenamefont {Isaeva}}]{Zeugner2019}%
  \BibitemOpen
  \bibfield  {author} {\bibinfo {author} {\bibfnamefont {A.}~\bibnamefont
  {Zeugner}}, \bibinfo {author} {\bibfnamefont {F.}~\bibnamefont {Nietschke}},
  \bibinfo {author} {\bibfnamefont {A.~U.}\ \bibnamefont {Wolter}}, \bibinfo
  {author} {\bibfnamefont {S.}~\bibnamefont {Ga{\ss}}}, \bibinfo {author}
  {\bibfnamefont {R.~C.}\ \bibnamefont {Vidal}}, \bibinfo {author}
  {\bibfnamefont {T.~R.}\ \bibnamefont {Peixoto}}, \bibinfo {author}
  {\bibfnamefont {D.}~\bibnamefont {Pohl}}, \bibinfo {author} {\bibfnamefont
  {C.}~\bibnamefont {Damm}}, \bibinfo {author} {\bibfnamefont {A.}~\bibnamefont
  {Lubk}}, \bibinfo {author} {\bibfnamefont {R.}~\bibnamefont {Hentrich}},
  \bibinfo {author} {\bibfnamefont {S.~K.}\ \bibnamefont {Moser}}, \bibinfo
  {author} {\bibfnamefont {C.}~\bibnamefont {Fornari}}, \bibinfo {author}
  {\bibfnamefont {C.~H.}\ \bibnamefont {Min}}, \bibinfo {author} {\bibfnamefont
  {S.}~\bibnamefont {Schatz}}, \bibinfo {author} {\bibfnamefont
  {K.}~\bibnamefont {Ki{\ss}ner}}, \bibinfo {author} {\bibfnamefont
  {M.}~\bibnamefont {{\"{U}}nzelmann}}, \bibinfo {author} {\bibfnamefont
  {M.}~\bibnamefont {Kaiser}}, \bibinfo {author} {\bibfnamefont
  {F.}~\bibnamefont {Scaravaggi}}, \bibinfo {author} {\bibfnamefont
  {B.}~\bibnamefont {Rellinghaus}}, \bibinfo {author} {\bibfnamefont
  {K.}~\bibnamefont {Nielsch}}, \bibinfo {author} {\bibfnamefont
  {C.}~\bibnamefont {Hess}}, \bibinfo {author} {\bibfnamefont {B.}~\bibnamefont
  {B{\"{u}}chner}}, \bibinfo {author} {\bibfnamefont {F.}~\bibnamefont
  {Reinert}}, \bibinfo {author} {\bibfnamefont {H.}~\bibnamefont {Bentmann}},
  \bibinfo {author} {\bibfnamefont {O.}~\bibnamefont {Oeckler}}, \bibinfo
  {author} {\bibfnamefont {T.}~\bibnamefont {Doert}}, \bibinfo {author}
  {\bibfnamefont {M.}~\bibnamefont {Ruck}},\ and\ \bibinfo {author}
  {\bibfnamefont {A.}~\bibnamefont {Isaeva}},\ }\href
  {https://doi.org/10.1021/acs.chemmater.8b05017} {\bibfield  {journal}
  {\bibinfo  {journal} {Chem. Mater.}\ }\textbf {\bibinfo {volume} {31}},\
  \bibinfo {pages} {2795} (\bibinfo {year} {2019})}\BibitemShut {NoStop}%
\bibitem [{\citenamefont {Gong}\ \emph {et~al.}(2019)\citenamefont {Gong},
  \citenamefont {Guo}, \citenamefont {Li}, \citenamefont {Zhu}, \citenamefont
  {Liao}, \citenamefont {Liu}, \citenamefont {Zhang}, \citenamefont {Gu},
  \citenamefont {Tang}, \citenamefont {Feng}, \citenamefont {Zhang},
  \citenamefont {Li}, \citenamefont {Song}, \citenamefont {Wang}, \citenamefont
  {Yu}, \citenamefont {Chen}, \citenamefont {Wang}, \citenamefont {Yao},
  \citenamefont {Duan}, \citenamefont {Xu}, \citenamefont {Zhang},
  \citenamefont {Ma}, \citenamefont {Xue},\ and\ \citenamefont
  {He}}]{Gong2019}%
  \BibitemOpen
  \bibfield  {author} {\bibinfo {author} {\bibfnamefont {Y.}~\bibnamefont
  {Gong}}, \bibinfo {author} {\bibfnamefont {J.}~\bibnamefont {Guo}}, \bibinfo
  {author} {\bibfnamefont {J.}~\bibnamefont {Li}}, \bibinfo {author}
  {\bibfnamefont {K.}~\bibnamefont {Zhu}}, \bibinfo {author} {\bibfnamefont
  {M.}~\bibnamefont {Liao}}, \bibinfo {author} {\bibfnamefont {X.}~\bibnamefont
  {Liu}}, \bibinfo {author} {\bibfnamefont {Q.}~\bibnamefont {Zhang}}, \bibinfo
  {author} {\bibfnamefont {L.}~\bibnamefont {Gu}}, \bibinfo {author}
  {\bibfnamefont {L.}~\bibnamefont {Tang}}, \bibinfo {author} {\bibfnamefont
  {X.}~\bibnamefont {Feng}}, \bibinfo {author} {\bibfnamefont {D.}~\bibnamefont
  {Zhang}}, \bibinfo {author} {\bibfnamefont {W.}~\bibnamefont {Li}}, \bibinfo
  {author} {\bibfnamefont {C.}~\bibnamefont {Song}}, \bibinfo {author}
  {\bibfnamefont {L.}~\bibnamefont {Wang}}, \bibinfo {author} {\bibfnamefont
  {P.}~\bibnamefont {Yu}}, \bibinfo {author} {\bibfnamefont {X.}~\bibnamefont
  {Chen}}, \bibinfo {author} {\bibfnamefont {Y.}~\bibnamefont {Wang}}, \bibinfo
  {author} {\bibfnamefont {H.}~\bibnamefont {Yao}}, \bibinfo {author}
  {\bibfnamefont {W.}~\bibnamefont {Duan}}, \bibinfo {author} {\bibfnamefont
  {Y.}~\bibnamefont {Xu}}, \bibinfo {author} {\bibfnamefont {S.-C.}\
  \bibnamefont {Zhang}}, \bibinfo {author} {\bibfnamefont {X.}~\bibnamefont
  {Ma}}, \bibinfo {author} {\bibfnamefont {Q.-K.}\ \bibnamefont {Xue}},\ and\
  \bibinfo {author} {\bibfnamefont {K.}~\bibnamefont {He}},\ }\href
  {https://doi.org/10.1088/0256-307X/36/7/076801} {\bibfield  {journal}
  {\bibinfo  {journal} {Chin. Phys. Lett.}\ }\textbf {\bibinfo {volume} {36}},\
  \bibinfo {pages} {076801} (\bibinfo {year} {2019})}\BibitemShut {NoStop}%
\bibitem [{\citenamefont {Hasan}\ and\ \citenamefont {Kane}(2010)}]{Hasan2010}%
  \BibitemOpen
  \bibfield  {author} {\bibinfo {author} {\bibfnamefont {M.~Z.}\ \bibnamefont
  {Hasan}}\ and\ \bibinfo {author} {\bibfnamefont {C.~L.}\ \bibnamefont
  {Kane}},\ }\href {https://doi.org/10.1103/RevModPhys.82.3045} {\bibfield
  {journal} {\bibinfo  {journal} {Rev. Mod. Phys.}\ }\textbf {\bibinfo {volume}
  {82}},\ \bibinfo {pages} {3045} (\bibinfo {year} {2010})}\BibitemShut
  {NoStop}%
\bibitem [{\citenamefont {Li}\ \emph {et~al.}(2019)\citenamefont {Li},
  \citenamefont {Li}, \citenamefont {Du}, \citenamefont {Wang}, \citenamefont
  {Gu}, \citenamefont {Zhang}, \citenamefont {He}, \citenamefont {Duan},\ and\
  \citenamefont {Xu}}]{Li2019}%
  \BibitemOpen
  \bibfield  {author} {\bibinfo {author} {\bibfnamefont {J.}~\bibnamefont
  {Li}}, \bibinfo {author} {\bibfnamefont {Y.}~\bibnamefont {Li}}, \bibinfo
  {author} {\bibfnamefont {S.}~\bibnamefont {Du}}, \bibinfo {author}
  {\bibfnamefont {Z.}~\bibnamefont {Wang}}, \bibinfo {author} {\bibfnamefont
  {B.-L.}\ \bibnamefont {Gu}}, \bibinfo {author} {\bibfnamefont {S.-C.}\
  \bibnamefont {Zhang}}, \bibinfo {author} {\bibfnamefont {K.}~\bibnamefont
  {He}}, \bibinfo {author} {\bibfnamefont {W.}~\bibnamefont {Duan}},\ and\
  \bibinfo {author} {\bibfnamefont {Y.}~\bibnamefont {Xu}},\ }\href
  {https://doi.org/10.1126/sciadv.aaw5685} {\bibfield  {journal} {\bibinfo
  {journal} {Sci. Adv.}\ }\textbf {\bibinfo {volume} {5}},\ \bibinfo {pages}
  {eaaw5685} (\bibinfo {year} {2019})}\BibitemShut {NoStop}%
\bibitem [{\citenamefont {Zhang}\ \emph {et~al.}(2019)\citenamefont {Zhang},
  \citenamefont {Shi}, \citenamefont {Zhu}, \citenamefont {Xing}, \citenamefont
  {Zhang},\ and\ \citenamefont {Wang}}]{Zhang2019}%
  \BibitemOpen
  \bibfield  {author} {\bibinfo {author} {\bibfnamefont {D.}~\bibnamefont
  {Zhang}}, \bibinfo {author} {\bibfnamefont {M.}~\bibnamefont {Shi}}, \bibinfo
  {author} {\bibfnamefont {T.}~\bibnamefont {Zhu}}, \bibinfo {author}
  {\bibfnamefont {D.}~\bibnamefont {Xing}}, \bibinfo {author} {\bibfnamefont
  {H.}~\bibnamefont {Zhang}},\ and\ \bibinfo {author} {\bibfnamefont
  {J.}~\bibnamefont {Wang}},\ }\href
  {https://doi.org/10.1103/PhysRevLett.122.206401} {\bibfield  {journal}
  {\bibinfo  {journal} {Phys. Rev. Lett.}\ }\textbf {\bibinfo {volume} {122}},\
  \bibinfo {pages} {206401} (\bibinfo {year} {2019})}\BibitemShut {NoStop}%
\bibitem [{\citenamefont {Otrokov}\ \emph
  {et~al.}(2019{\natexlab{b}})\citenamefont {Otrokov}, \citenamefont {Rusinov},
  \citenamefont {Blanco-Rey}, \citenamefont {Hoffmann}, \citenamefont
  {Vyazovskaya}, \citenamefont {Eremeev}, \citenamefont {Ernst}, \citenamefont
  {Echenique}, \citenamefont {Arnau},\ and\ \citenamefont
  {Chulkov}}]{Otrokov2019Cal}%
  \BibitemOpen
  \bibfield  {author} {\bibinfo {author} {\bibfnamefont {M.~M.}\ \bibnamefont
  {Otrokov}}, \bibinfo {author} {\bibfnamefont {I.~P.}\ \bibnamefont
  {Rusinov}}, \bibinfo {author} {\bibfnamefont {M.}~\bibnamefont {Blanco-Rey}},
  \bibinfo {author} {\bibfnamefont {M.}~\bibnamefont {Hoffmann}}, \bibinfo
  {author} {\bibfnamefont {A.~Y.}\ \bibnamefont {Vyazovskaya}}, \bibinfo
  {author} {\bibfnamefont {S.~V.}\ \bibnamefont {Eremeev}}, \bibinfo {author}
  {\bibfnamefont {A.}~\bibnamefont {Ernst}}, \bibinfo {author} {\bibfnamefont
  {P.~M.}\ \bibnamefont {Echenique}}, \bibinfo {author} {\bibfnamefont
  {A.}~\bibnamefont {Arnau}},\ and\ \bibinfo {author} {\bibfnamefont {E.~V.}\
  \bibnamefont {Chulkov}},\ }\href
  {https://doi.org/10.1103/PhysRevLett.122.107202} {\bibfield  {journal}
  {\bibinfo  {journal} {Phys. Rev. Lett.}\ }\textbf {\bibinfo {volume} {122}},\
  \bibinfo {pages} {107202} (\bibinfo {year} {2019}{\natexlab{b}})}\BibitemShut
  {NoStop}%
\bibitem [{\citenamefont {Yan}\ \emph {et~al.}(2019)\citenamefont {Yan},
  \citenamefont {Zhang}, \citenamefont {Heitmann}, \citenamefont {Huang},
  \citenamefont {Chen}, \citenamefont {Cheng}, \citenamefont {Wu},
  \citenamefont {Vaknin}, \citenamefont {Sales},\ and\ \citenamefont
  {McQueeney}}]{Yan2019}%
  \BibitemOpen
  \bibfield  {author} {\bibinfo {author} {\bibfnamefont {J.-Q.}\ \bibnamefont
  {Yan}}, \bibinfo {author} {\bibfnamefont {Q.}~\bibnamefont {Zhang}}, \bibinfo
  {author} {\bibfnamefont {T.}~\bibnamefont {Heitmann}}, \bibinfo {author}
  {\bibfnamefont {Z.}~\bibnamefont {Huang}}, \bibinfo {author} {\bibfnamefont
  {K.~Y.}\ \bibnamefont {Chen}}, \bibinfo {author} {\bibfnamefont {J.-G.}\
  \bibnamefont {Cheng}}, \bibinfo {author} {\bibfnamefont {W.}~\bibnamefont
  {Wu}}, \bibinfo {author} {\bibfnamefont {D.}~\bibnamefont {Vaknin}}, \bibinfo
  {author} {\bibfnamefont {B.~C.}\ \bibnamefont {Sales}},\ and\ \bibinfo
  {author} {\bibfnamefont {R.~J.}\ \bibnamefont {McQueeney}},\ }\href
  {https://doi.org/10.1103/PhysRevMaterials.3.064202} {\bibfield  {journal}
  {\bibinfo  {journal} {Phys. Rev. Mater.}\ }\textbf {\bibinfo {volume} {3}},\
  \bibinfo {pages} {064202} (\bibinfo {year} {2019})}\BibitemShut {NoStop}%
\bibitem [{\citenamefont {Mogi}\ \emph {et~al.}(2017)\citenamefont {Mogi},
  \citenamefont {Kawamura}, \citenamefont {Yoshimi}, \citenamefont {Tsukazaki},
  \citenamefont {Kozuka}, \citenamefont {Shirakawa}, \citenamefont {Takahashi},
  \citenamefont {Kawasaki},\ and\ \citenamefont {Tokura}}]{Mogi2017}%
  \BibitemOpen
  \bibfield  {author} {\bibinfo {author} {\bibfnamefont {M.}~\bibnamefont
  {Mogi}}, \bibinfo {author} {\bibfnamefont {M.}~\bibnamefont {Kawamura}},
  \bibinfo {author} {\bibfnamefont {R.}~\bibnamefont {Yoshimi}}, \bibinfo
  {author} {\bibfnamefont {A.}~\bibnamefont {Tsukazaki}}, \bibinfo {author}
  {\bibfnamefont {Y.}~\bibnamefont {Kozuka}}, \bibinfo {author} {\bibfnamefont
  {N.}~\bibnamefont {Shirakawa}}, \bibinfo {author} {\bibfnamefont {K.~S.}\
  \bibnamefont {Takahashi}}, \bibinfo {author} {\bibfnamefont {M.}~\bibnamefont
  {Kawasaki}},\ and\ \bibinfo {author} {\bibfnamefont {Y.}~\bibnamefont
  {Tokura}},\ }\href {https://doi.org/10.1038/nmat4855} {\bibfield  {journal}
  {\bibinfo  {journal} {Nat. Mater.}\ }\textbf {\bibinfo {volume} {16}},\
  \bibinfo {pages} {516} (\bibinfo {year} {2017})}\BibitemShut {NoStop}%
\bibitem [{\citenamefont {Padmanabhan}\ \emph {et~al.}(2022)\citenamefont
  {Padmanabhan}, \citenamefont {Poore}, \citenamefont {Kim}, \citenamefont
  {Koocher}, \citenamefont {Stoica}, \citenamefont {Puggioni}, \citenamefont
  {Wang}, \citenamefont {Shen}, \citenamefont {Reid}, \citenamefont {Gu},
  \citenamefont {Wetherington}, \citenamefont {Lee}, \citenamefont {Schaller},
  \citenamefont {Mao}, \citenamefont {Lindenberg}, \citenamefont {Wang},
  \citenamefont {Rondinelli}, \citenamefont {Averitt},\ and\ \citenamefont
  {Gopalan}}]{Padmanabhan2022}%
  \BibitemOpen
  \bibfield  {author} {\bibinfo {author} {\bibfnamefont {H.}~\bibnamefont
  {Padmanabhan}}, \bibinfo {author} {\bibfnamefont {M.}~\bibnamefont {Poore}},
  \bibinfo {author} {\bibfnamefont {P.~K.}\ \bibnamefont {Kim}}, \bibinfo
  {author} {\bibfnamefont {N.~Z.}\ \bibnamefont {Koocher}}, \bibinfo {author}
  {\bibfnamefont {V.~A.}\ \bibnamefont {Stoica}}, \bibinfo {author}
  {\bibfnamefont {D.}~\bibnamefont {Puggioni}}, \bibinfo {author}
  {\bibfnamefont {H.}~\bibnamefont {Wang}}, \bibinfo {author} {\bibfnamefont
  {X.}~\bibnamefont {Shen}}, \bibinfo {author} {\bibfnamefont {A.~H.}\
  \bibnamefont {Reid}}, \bibinfo {author} {\bibfnamefont {M.}~\bibnamefont
  {Gu}}, \bibinfo {author} {\bibfnamefont {M.}~\bibnamefont {Wetherington}},
  \bibinfo {author} {\bibfnamefont {S.~H.}\ \bibnamefont {Lee}}, \bibinfo
  {author} {\bibfnamefont {R.~D.}\ \bibnamefont {Schaller}}, \bibinfo {author}
  {\bibfnamefont {Z.}~\bibnamefont {Mao}}, \bibinfo {author} {\bibfnamefont
  {A.~M.}\ \bibnamefont {Lindenberg}}, \bibinfo {author} {\bibfnamefont
  {X.}~\bibnamefont {Wang}}, \bibinfo {author} {\bibfnamefont {J.~M.}\
  \bibnamefont {Rondinelli}}, \bibinfo {author} {\bibfnamefont {R.~D.}\
  \bibnamefont {Averitt}},\ and\ \bibinfo {author} {\bibfnamefont
  {V.}~\bibnamefont {Gopalan}},\ }\href
  {https://doi.org/10.1038/s41467-022-29545-5} {\bibfield  {journal} {\bibinfo
  {journal} {Nat. Commun.}\ }\textbf {\bibinfo {volume} {13}},\ \bibinfo
  {pages} {1929} (\bibinfo {year} {2022})}\BibitemShut {NoStop}%
\bibitem [{\citenamefont {Choe}\ \emph {et~al.}(2021)\citenamefont {Choe},
  \citenamefont {Lujan}, \citenamefont {Rodriguez-Vega}, \citenamefont {Ye},
  \citenamefont {Leonardo}, \citenamefont {Quan}, \citenamefont {Nunley},
  \citenamefont {Chang}, \citenamefont {Lee}, \citenamefont {Yan},
  \citenamefont {Fiete}, \citenamefont {He},\ and\ \citenamefont
  {Li}}]{doi:10.1021/acs.nanolett.1c01719}%
  \BibitemOpen
  \bibfield  {author} {\bibinfo {author} {\bibfnamefont {J.}~\bibnamefont
  {Choe}}, \bibinfo {author} {\bibfnamefont {D.}~\bibnamefont {Lujan}},
  \bibinfo {author} {\bibfnamefont {M.}~\bibnamefont {Rodriguez-Vega}},
  \bibinfo {author} {\bibfnamefont {Z.}~\bibnamefont {Ye}}, \bibinfo {author}
  {\bibfnamefont {A.}~\bibnamefont {Leonardo}}, \bibinfo {author}
  {\bibfnamefont {J.}~\bibnamefont {Quan}}, \bibinfo {author} {\bibfnamefont
  {T.~N.}\ \bibnamefont {Nunley}}, \bibinfo {author} {\bibfnamefont {L.-J.}\
  \bibnamefont {Chang}}, \bibinfo {author} {\bibfnamefont {S.-F.}\ \bibnamefont
  {Lee}}, \bibinfo {author} {\bibfnamefont {J.}~\bibnamefont {Yan}}, \bibinfo
  {author} {\bibfnamefont {G.~A.}\ \bibnamefont {Fiete}}, \bibinfo {author}
  {\bibfnamefont {R.}~\bibnamefont {He}},\ and\ \bibinfo {author}
  {\bibfnamefont {X.}~\bibnamefont {Li}},\ }\href
  {https://doi.org/10.1021/acs.nanolett.1c01719} {\bibfield  {journal}
  {\bibinfo  {journal} {Nano Lett.}\ }\textbf {\bibinfo {volume} {21}},\
  \bibinfo {pages} {6139} (\bibinfo {year} {2021})}\BibitemShut {NoStop}%
\bibitem [{\citenamefont {Bartram}\ \emph {et~al.}(2022)\citenamefont
  {Bartram}, \citenamefont {Leng}, \citenamefont {Wang}, \citenamefont {Liu},
  \citenamefont {Chen}, \citenamefont {Peng}, \citenamefont {Li}, \citenamefont
  {Yu}, \citenamefont {Wu}, \citenamefont {Lin}, \citenamefont {Zhang},
  \citenamefont {Tan},\ and\ \citenamefont {Yang}}]{Bartram2022}%
  \BibitemOpen
  \bibfield  {author} {\bibinfo {author} {\bibfnamefont {F.~M.}\ \bibnamefont
  {Bartram}}, \bibinfo {author} {\bibfnamefont {Y.-C.}\ \bibnamefont {Leng}},
  \bibinfo {author} {\bibfnamefont {Y.}~\bibnamefont {Wang}}, \bibinfo {author}
  {\bibfnamefont {L.}~\bibnamefont {Liu}}, \bibinfo {author} {\bibfnamefont
  {X.}~\bibnamefont {Chen}}, \bibinfo {author} {\bibfnamefont {H.}~\bibnamefont
  {Peng}}, \bibinfo {author} {\bibfnamefont {H.}~\bibnamefont {Li}}, \bibinfo
  {author} {\bibfnamefont {P.}~\bibnamefont {Yu}}, \bibinfo {author}
  {\bibfnamefont {Y.}~\bibnamefont {Wu}}, \bibinfo {author} {\bibfnamefont
  {M.-L.}\ \bibnamefont {Lin}}, \bibinfo {author} {\bibfnamefont
  {J.}~\bibnamefont {Zhang}}, \bibinfo {author} {\bibfnamefont {P.-H.}\
  \bibnamefont {Tan}},\ and\ \bibinfo {author} {\bibfnamefont {L.}~\bibnamefont
  {Yang}},\ }\href {https://doi.org/10.1038/s41535-022-00495-x} {\bibfield
  {journal} {\bibinfo  {journal} {npj Quantum Mater.}\ }\textbf {\bibinfo
  {volume} {7}},\ \bibinfo {pages} {84} (\bibinfo {year} {2022})}\BibitemShut
  {NoStop}%
\bibitem [{\citenamefont {Qi}\ \emph {et~al.}(2010)\citenamefont {Qi},
  \citenamefont {Chen}, \citenamefont {Yu}, \citenamefont {Cadden-Zimansky},
  \citenamefont {Smirnov}, \citenamefont {Tolk}, \citenamefont {Miotkowski},
  \citenamefont {Cao}, \citenamefont {Chen}, \citenamefont {Wu}, \citenamefont
  {Qiao},\ and\ \citenamefont {Jiang}}]{Qi2010}%
  \BibitemOpen
  \bibfield  {author} {\bibinfo {author} {\bibfnamefont {J.}~\bibnamefont
  {Qi}}, \bibinfo {author} {\bibfnamefont {X.}~\bibnamefont {Chen}}, \bibinfo
  {author} {\bibfnamefont {W.}~\bibnamefont {Yu}}, \bibinfo {author}
  {\bibfnamefont {P.}~\bibnamefont {Cadden-Zimansky}}, \bibinfo {author}
  {\bibfnamefont {D.}~\bibnamefont {Smirnov}}, \bibinfo {author} {\bibfnamefont
  {N.~H.}\ \bibnamefont {Tolk}}, \bibinfo {author} {\bibfnamefont
  {I.}~\bibnamefont {Miotkowski}}, \bibinfo {author} {\bibfnamefont
  {H.}~\bibnamefont {Cao}}, \bibinfo {author} {\bibfnamefont {Y.~P.}\
  \bibnamefont {Chen}}, \bibinfo {author} {\bibfnamefont {Y.}~\bibnamefont
  {Wu}}, \bibinfo {author} {\bibfnamefont {S.}~\bibnamefont {Qiao}},\ and\
  \bibinfo {author} {\bibfnamefont {Z.}~\bibnamefont {Jiang}},\ }\href
  {https://doi.org/10.1063/1.3513826} {\bibfield  {journal} {\bibinfo
  {journal} {Appl. Phys. Lett.}\ }\textbf {\bibinfo {volume} {97}},\ \bibinfo
  {pages} {182102} (\bibinfo {year} {2010})}\BibitemShut {NoStop}%
\bibitem [{\citenamefont {Lai}\ \emph {et~al.}(2014)\citenamefont {Lai},
  \citenamefont {Chen}, \citenamefont {Wu},\ and\ \citenamefont
  {Liu}}]{Lai2014}%
  \BibitemOpen
  \bibfield  {author} {\bibinfo {author} {\bibfnamefont {Y.-P.}\ \bibnamefont
  {Lai}}, \bibinfo {author} {\bibfnamefont {H.-J.}\ \bibnamefont {Chen}},
  \bibinfo {author} {\bibfnamefont {K.-H.}\ \bibnamefont {Wu}},\ and\ \bibinfo
  {author} {\bibfnamefont {J.-M.}\ \bibnamefont {Liu}},\ }\href
  {https://doi.org/10.1063/1.4904009} {\bibfield  {journal} {\bibinfo
  {journal} {Appl. Phys. Lett.}\ }\textbf {\bibinfo {volume} {105}},\ \bibinfo
  {pages} {232110} (\bibinfo {year} {2014})}\BibitemShut {NoStop}%
\bibitem [{\citenamefont {Cheng}\ \emph {et~al.}(2014)\citenamefont {Cheng},
  \citenamefont {La-o vorakiat}, \citenamefont {Tang}, \citenamefont {Nair},
  \citenamefont {Xia}, \citenamefont {Wang}, \citenamefont {Zhu},\ and\
  \citenamefont {Chia}}]{Cheng2014}%
  \BibitemOpen
  \bibfield  {author} {\bibinfo {author} {\bibfnamefont {L.}~\bibnamefont
  {Cheng}}, \bibinfo {author} {\bibfnamefont {C.}~\bibnamefont {La-o
  vorakiat}}, \bibinfo {author} {\bibfnamefont {C.~S.}\ \bibnamefont {Tang}},
  \bibinfo {author} {\bibfnamefont {S.~K.}\ \bibnamefont {Nair}}, \bibinfo
  {author} {\bibfnamefont {B.}~\bibnamefont {Xia}}, \bibinfo {author}
  {\bibfnamefont {L.}~\bibnamefont {Wang}}, \bibinfo {author} {\bibfnamefont
  {J.-X.}\ \bibnamefont {Zhu}},\ and\ \bibinfo {author} {\bibfnamefont
  {E.~E.~M.}\ \bibnamefont {Chia}},\ }\href {https://doi.org/10.1063/1.4879831}
  {\bibfield  {journal} {\bibinfo  {journal} {Appl. Phys. Lett.}\ }\textbf
  {\bibinfo {volume} {104}},\ \bibinfo {pages} {211906} (\bibinfo {year}
  {2014})}\BibitemShut {NoStop}%
\bibitem [{\citenamefont {Wang}\ \emph {et~al.}(2016)\citenamefont {Wang},
  \citenamefont {Qiao}, \citenamefont {Jiang}, \citenamefont {Luo},\ and\
  \citenamefont {Qi}}]{Wang2015}%
  \BibitemOpen
  \bibfield  {author} {\bibinfo {author} {\bibfnamefont {M.~C.}\ \bibnamefont
  {Wang}}, \bibinfo {author} {\bibfnamefont {S.}~\bibnamefont {Qiao}}, \bibinfo
  {author} {\bibfnamefont {Z.}~\bibnamefont {Jiang}}, \bibinfo {author}
  {\bibfnamefont {S.~N.}\ \bibnamefont {Luo}},\ and\ \bibinfo {author}
  {\bibfnamefont {J.}~\bibnamefont {Qi}},\ }\href
  {https://doi.org/10.1103/PhysRevLett.116.036601} {\bibfield  {journal}
  {\bibinfo  {journal} {Phys. Rev. Lett.}\ }\textbf {\bibinfo {volume} {116}},\
  \bibinfo {pages} {036601} (\bibinfo {year} {2016})}\BibitemShut {NoStop}%
\bibitem [{\citenamefont {Hilton}\ and\ \citenamefont
  {Tang}(2002)}]{Hilton2002}%
  \BibitemOpen
  \bibfield  {author} {\bibinfo {author} {\bibfnamefont {D.~J.}\ \bibnamefont
  {Hilton}}\ and\ \bibinfo {author} {\bibfnamefont {C.~L.}\ \bibnamefont
  {Tang}},\ }\href {https://doi.org/10.1103/PhysRevLett.89.146601} {\bibfield
  {journal} {\bibinfo  {journal} {Phys. Rev. Lett.}\ }\textbf {\bibinfo
  {volume} {89}},\ \bibinfo {pages} {146601} (\bibinfo {year}
  {2002})}\BibitemShut {NoStop}%
\bibitem [{\citenamefont {Dai}\ \emph {et~al.}(2015)\citenamefont {Dai},
  \citenamefont {Bowlan}, \citenamefont {Li}, \citenamefont {Miao},
  \citenamefont {Wu}, \citenamefont {Kong}, \citenamefont {Richard},
  \citenamefont {Shi}, \citenamefont {Trugman}, \citenamefont {Zhu},
  \citenamefont {Ding}, \citenamefont {Taylor}, \citenamefont {Yarotski},\ and\
  \citenamefont {Prasankumar}}]{Dai2015}%
  \BibitemOpen
  \bibfield  {author} {\bibinfo {author} {\bibfnamefont {Y.~M.}\ \bibnamefont
  {Dai}}, \bibinfo {author} {\bibfnamefont {J.}~\bibnamefont {Bowlan}},
  \bibinfo {author} {\bibfnamefont {H.}~\bibnamefont {Li}}, \bibinfo {author}
  {\bibfnamefont {H.}~\bibnamefont {Miao}}, \bibinfo {author} {\bibfnamefont
  {S.~F.}\ \bibnamefont {Wu}}, \bibinfo {author} {\bibfnamefont {W.~D.}\
  \bibnamefont {Kong}}, \bibinfo {author} {\bibfnamefont {P.}~\bibnamefont
  {Richard}}, \bibinfo {author} {\bibfnamefont {Y.~G.}\ \bibnamefont {Shi}},
  \bibinfo {author} {\bibfnamefont {S.~A.}\ \bibnamefont {Trugman}}, \bibinfo
  {author} {\bibfnamefont {J.-X.}\ \bibnamefont {Zhu}}, \bibinfo {author}
  {\bibfnamefont {H.}~\bibnamefont {Ding}}, \bibinfo {author} {\bibfnamefont
  {A.~J.}\ \bibnamefont {Taylor}}, \bibinfo {author} {\bibfnamefont {D.~A.}\
  \bibnamefont {Yarotski}},\ and\ \bibinfo {author} {\bibfnamefont {R.~P.}\
  \bibnamefont {Prasankumar}},\ }\href
  {https://doi.org/10.1103/PhysRevB.92.161104} {\bibfield  {journal} {\bibinfo
  {journal} {Phys. Rev. B}\ }\textbf {\bibinfo {volume} {92}},\ \bibinfo
  {pages} {161104(R)} (\bibinfo {year} {2015})}\BibitemShut {NoStop}%
\bibitem [{\citenamefont {Cheng}\ \emph {et~al.}(2022)\citenamefont {Cheng},
  \citenamefont {Fei}, \citenamefont {Hu}, \citenamefont {Dai}, \citenamefont
  {Song},\ and\ \citenamefont {Qi}}]{Cheng2022}%
  \BibitemOpen
  \bibfield  {author} {\bibinfo {author} {\bibfnamefont {L.}~\bibnamefont
  {Cheng}}, \bibinfo {author} {\bibfnamefont {F.~C.}\ \bibnamefont {Fei}},
  \bibinfo {author} {\bibfnamefont {H.}~\bibnamefont {Hu}}, \bibinfo {author}
  {\bibfnamefont {Y.~M.}\ \bibnamefont {Dai}}, \bibinfo {author} {\bibfnamefont
  {F.~Q.}\ \bibnamefont {Song}},\ and\ \bibinfo {author} {\bibfnamefont
  {J.}~\bibnamefont {Qi}},\ }\href
  {https://doi.org/10.1103/PhysRevB.106.104308} {\bibfield  {journal} {\bibinfo
   {journal} {Phys. Rev. B}\ }\textbf {\bibinfo {volume} {106}},\ \bibinfo
  {pages} {104308} (\bibinfo {year} {2022})}\BibitemShut {NoStop}%
\bibitem [{\citenamefont {Thielemann-K\"uhn}\ \emph {et~al.}(2017)\citenamefont
  {Thielemann-K\"uhn}, \citenamefont {Schick}, \citenamefont {Pontius},
  \citenamefont {Trabant}, \citenamefont {Mitzner}, \citenamefont {Holldack},
  \citenamefont {Zabel}, \citenamefont {F\"ohlisch},\ and\ \citenamefont
  {Sch\"u\ss{}ler-Langeheine}}]{Thielemann-Kuhn2017a}%
  \BibitemOpen
  \bibfield  {author} {\bibinfo {author} {\bibfnamefont {N.}~\bibnamefont
  {Thielemann-K\"uhn}}, \bibinfo {author} {\bibfnamefont {D.}~\bibnamefont
  {Schick}}, \bibinfo {author} {\bibfnamefont {N.}~\bibnamefont {Pontius}},
  \bibinfo {author} {\bibfnamefont {C.}~\bibnamefont {Trabant}}, \bibinfo
  {author} {\bibfnamefont {R.}~\bibnamefont {Mitzner}}, \bibinfo {author}
  {\bibfnamefont {K.}~\bibnamefont {Holldack}}, \bibinfo {author}
  {\bibfnamefont {H.}~\bibnamefont {Zabel}}, \bibinfo {author} {\bibfnamefont
  {A.}~\bibnamefont {F\"ohlisch}},\ and\ \bibinfo {author} {\bibfnamefont
  {C.}~\bibnamefont {Sch\"u\ss{}ler-Langeheine}},\ }\href
  {https://doi.org/10.1103/PhysRevLett.119.197202} {\bibfield  {journal}
  {\bibinfo  {journal} {Phys. Rev. Lett.}\ }\textbf {\bibinfo {volume} {119}},\
  \bibinfo {pages} {197202} (\bibinfo {year} {2017})}\BibitemShut {NoStop}%
\bibitem [{\citenamefont {Rettig}\ \emph {et~al.}(2012)\citenamefont {Rettig},
  \citenamefont {Cort\'es}, \citenamefont {Thirupathaiah}, \citenamefont
  {Gegenwart}, \citenamefont {Jeevan}, \citenamefont {Wolf}, \citenamefont
  {Fink},\ and\ \citenamefont {Bovensiepen}}]{Rettig2012}%
  \BibitemOpen
  \bibfield  {author} {\bibinfo {author} {\bibfnamefont {L.}~\bibnamefont
  {Rettig}}, \bibinfo {author} {\bibfnamefont {R.}~\bibnamefont {Cort\'es}},
  \bibinfo {author} {\bibfnamefont {S.}~\bibnamefont {Thirupathaiah}}, \bibinfo
  {author} {\bibfnamefont {P.}~\bibnamefont {Gegenwart}}, \bibinfo {author}
  {\bibfnamefont {H.~S.}\ \bibnamefont {Jeevan}}, \bibinfo {author}
  {\bibfnamefont {M.}~\bibnamefont {Wolf}}, \bibinfo {author} {\bibfnamefont
  {J.}~\bibnamefont {Fink}},\ and\ \bibinfo {author} {\bibfnamefont
  {U.}~\bibnamefont {Bovensiepen}},\ }\href
  {https://doi.org/10.1103/PhysRevLett.108.097002} {\bibfield  {journal}
  {\bibinfo  {journal} {Phys. Rev. Lett.}\ }\textbf {\bibinfo {volume} {108}},\
  \bibinfo {pages} {097002} (\bibinfo {year} {2012})}\BibitemShut {NoStop}%
\bibitem [{\citenamefont {Carpene}\ \emph {et~al.}(2008)\citenamefont
  {Carpene}, \citenamefont {Mancini}, \citenamefont {Dallera}, \citenamefont
  {Brenna}, \citenamefont {Puppin},\ and\ \citenamefont
  {De~Silvestri}}]{PhysRevB.78.174422}%
  \BibitemOpen
  \bibfield  {author} {\bibinfo {author} {\bibfnamefont {E.}~\bibnamefont
  {Carpene}}, \bibinfo {author} {\bibfnamefont {E.}~\bibnamefont {Mancini}},
  \bibinfo {author} {\bibfnamefont {C.}~\bibnamefont {Dallera}}, \bibinfo
  {author} {\bibfnamefont {M.}~\bibnamefont {Brenna}}, \bibinfo {author}
  {\bibfnamefont {E.}~\bibnamefont {Puppin}},\ and\ \bibinfo {author}
  {\bibfnamefont {S.}~\bibnamefont {De~Silvestri}},\ }\href
  {https://doi.org/10.1103/PhysRevB.78.174422} {\bibfield  {journal} {\bibinfo
  {journal} {Phys. Rev. B}\ }\textbf {\bibinfo {volume} {78}},\ \bibinfo
  {pages} {174422} (\bibinfo {year} {2008})}\BibitemShut {NoStop}%
\bibitem [{\citenamefont {Groeneveld}\ \emph {et~al.}(1995)\citenamefont
  {Groeneveld}, \citenamefont {Sprik},\ and\ \citenamefont
  {Lagendijk}}]{groeneveld1995}%
  \BibitemOpen
  \bibfield  {author} {\bibinfo {author} {\bibfnamefont {R.~H.~M.}\
  \bibnamefont {Groeneveld}}, \bibinfo {author} {\bibfnamefont
  {R.}~\bibnamefont {Sprik}},\ and\ \bibinfo {author} {\bibfnamefont
  {A.}~\bibnamefont {Lagendijk}},\ }\href
  {https://doi.org/10.1103/PhysRevB.51.11433} {\bibfield  {journal} {\bibinfo
  {journal} {Physical Review B}\ }\textbf {\bibinfo {volume} {51}},\ \bibinfo
  {pages} {11433} (\bibinfo {year} {1995})}\BibitemShut {NoStop}%
\bibitem [{\citenamefont {Schroder}(2005)}]{Schroder2005}%
  \BibitemOpen
  \bibfield  {author} {\bibinfo {author} {\bibfnamefont {D.~K.}\ \bibnamefont
  {Schroder}},\ }\href@noop {} { {\bibinfo {title} {Semiconductor Material
  and Device Characterization}}}\ (\bibinfo  {publisher} {JohnWiley and Sons
  Inc.},\ \bibinfo {year} {2005})\BibitemShut {NoStop}%
\bibitem [{\citenamefont {Sheu}\ \emph {et~al.}(2013)\citenamefont {Sheu},
  \citenamefont {Chien}, \citenamefont {Uher}, \citenamefont {Fahy},\ and\
  \citenamefont {Reis}}]{Sheu2013}%
  \BibitemOpen
  \bibfield  {author} {\bibinfo {author} {\bibfnamefont {Y.~M.}\ \bibnamefont
  {Sheu}}, \bibinfo {author} {\bibfnamefont {Y.~J.}\ \bibnamefont {Chien}},
  \bibinfo {author} {\bibfnamefont {C.}~\bibnamefont {Uher}}, \bibinfo {author}
  {\bibfnamefont {S.}~\bibnamefont {Fahy}},\ and\ \bibinfo {author}
  {\bibfnamefont {D.~A.}\ \bibnamefont {Reis}},\ }\href
  {https://doi.org/10.1103/PhysRevB.87.075429} {\bibfield  {journal} {\bibinfo
  {journal} {Phys. Rev. B}\ }\textbf {\bibinfo {volume} {87}},\ \bibinfo
  {pages} {075429} (\bibinfo {year} {2013})}\BibitemShut {NoStop}%
\bibitem [{\citenamefont {Li}\ \emph {et~al.}(2021)\citenamefont {Li},
  \citenamefont {Yu}, \citenamefont {Wang},\ and\ \citenamefont
  {Luo}}]{Li2021}%
  \BibitemOpen
  \bibfield  {author} {\bibinfo {author} {\bibfnamefont {P.}~\bibnamefont
  {Li}}, \bibinfo {author} {\bibfnamefont {J.}~\bibnamefont {Yu}}, \bibinfo
  {author} {\bibfnamefont {Y.}~\bibnamefont {Wang}},\ and\ \bibinfo {author}
  {\bibfnamefont {W.}~\bibnamefont {Luo}},\ }\href
  {https://doi.org/10.1103/PhysRevB.103.155118} {\bibfield  {journal} {\bibinfo
   {journal} {Phys. Rev. B}\ }\textbf {\bibinfo {volume} {103}},\ \bibinfo
  {pages} {155118} (\bibinfo {year} {2021})}\BibitemShut {NoStop}%
\bibitem [{\citenamefont {Qi}\ \emph {et~al.}(2012)\citenamefont {Qi},
  \citenamefont {Yan}, \citenamefont {Zhou}, \citenamefont {Zhu}, \citenamefont
  {Trugman}, \citenamefont {Taylor}, \citenamefont {Jia},\ and\ \citenamefont
  {Prasankumar}}]{Qi2012}%
  \BibitemOpen
  \bibfield  {author} {\bibinfo {author} {\bibfnamefont {J.}~\bibnamefont
  {Qi}}, \bibinfo {author} {\bibfnamefont {L.}~\bibnamefont {Yan}}, \bibinfo
  {author} {\bibfnamefont {H.~D.}\ \bibnamefont {Zhou}}, \bibinfo {author}
  {\bibfnamefont {J.-X.}\ \bibnamefont {Zhu}}, \bibinfo {author} {\bibfnamefont
  {S.~A.}\ \bibnamefont {Trugman}}, \bibinfo {author} {\bibfnamefont {A.~J.}\
  \bibnamefont {Taylor}}, \bibinfo {author} {\bibfnamefont {Q.~X.}\
  \bibnamefont {Jia}},\ and\ \bibinfo {author} {\bibfnamefont {R.~P.}\
  \bibnamefont {Prasankumar}},\ }\href {https://doi.org/10.1063/1.4754294}
  {\bibfield  {journal} {\bibinfo  {journal} {Appl. Phys. Lett.}\ }\textbf
  {\bibinfo {volume} {101}},\ \bibinfo {pages} {122904} (\bibinfo {year}
  {2012})}\BibitemShut {NoStop}%
\bibitem [{\citenamefont {Vaterlaus}\ \emph {et~al.}(1991)\citenamefont
  {Vaterlaus}, \citenamefont {Beutler},\ and\ \citenamefont
  {Meier}}]{Vaterlaus1991}%
  \BibitemOpen
  \bibfield  {author} {\bibinfo {author} {\bibfnamefont {A.}~\bibnamefont
  {Vaterlaus}}, \bibinfo {author} {\bibfnamefont {T.}~\bibnamefont {Beutler}},\
  and\ \bibinfo {author} {\bibfnamefont {F.}~\bibnamefont {Meier}},\ }\href
  {https://doi.org/10.1103/PhysRevLett.67.3314} {\bibfield  {journal} {\bibinfo
   {journal} {Phys. Rev. Lett.}\ }\textbf {\bibinfo {volume} {67}},\ \bibinfo
  {pages} {3314} (\bibinfo {year} {1991})}\BibitemShut {NoStop}%
\bibitem [{\citenamefont {Yuan}\ \emph {et~al.}(2019)\citenamefont {Yuan},
  \citenamefont {Kissin}, \citenamefont {Puggioni}, \citenamefont {Cremin},
  \citenamefont {Lei}, \citenamefont {Wang}, \citenamefont {Mao}, \citenamefont
  {Rondinelli}, \citenamefont {Averitt},\ and\ \citenamefont
  {Gopalan}}]{Yuan2019}%
  \BibitemOpen
  \bibfield  {author} {\bibinfo {author} {\bibfnamefont {Y.}~\bibnamefont
  {Yuan}}, \bibinfo {author} {\bibfnamefont {P.}~\bibnamefont {Kissin}},
  \bibinfo {author} {\bibfnamefont {D.}~\bibnamefont {Puggioni}}, \bibinfo
  {author} {\bibfnamefont {K.}~\bibnamefont {Cremin}}, \bibinfo {author}
  {\bibfnamefont {S.}~\bibnamefont {Lei}}, \bibinfo {author} {\bibfnamefont
  {Y.}~\bibnamefont {Wang}}, \bibinfo {author} {\bibfnamefont {Z.}~\bibnamefont
  {Mao}}, \bibinfo {author} {\bibfnamefont {J.~M.}\ \bibnamefont {Rondinelli}},
  \bibinfo {author} {\bibfnamefont {R.~D.}\ \bibnamefont {Averitt}},\ and\
  \bibinfo {author} {\bibfnamefont {V.}~\bibnamefont {Gopalan}},\ }\href
  {https://doi.org/10.1103/PhysRevB.99.155111} {\bibfield  {journal} {\bibinfo
  {journal} {Phys. Rev. B}\ }\textbf {\bibinfo {volume} {99}},\ \bibinfo
  {pages} {155111} (\bibinfo {year} {2019})}\BibitemShut {NoStop}%
\bibitem [{\citenamefont {Chia}\ \emph {et~al.}(2007)\citenamefont {Chia},
  \citenamefont {Zhu}, \citenamefont {Talbayev}, \citenamefont {Averitt},
  \citenamefont {Taylor}, \citenamefont {Oh}, \citenamefont {Jo},\ and\
  \citenamefont {Lee}}]{Chia2007}%
  \BibitemOpen
  \bibfield  {author} {\bibinfo {author} {\bibfnamefont {E.~E.~M.}\
  \bibnamefont {Chia}}, \bibinfo {author} {\bibfnamefont {J.-X.}\ \bibnamefont
  {Zhu}}, \bibinfo {author} {\bibfnamefont {D.}~\bibnamefont {Talbayev}},
  \bibinfo {author} {\bibfnamefont {R.~D.}\ \bibnamefont {Averitt}}, \bibinfo
  {author} {\bibfnamefont {A.~J.}\ \bibnamefont {Taylor}}, \bibinfo {author}
  {\bibfnamefont {K.-H.}\ \bibnamefont {Oh}}, \bibinfo {author} {\bibfnamefont
  {I.-S.}\ \bibnamefont {Jo}},\ and\ \bibinfo {author} {\bibfnamefont {S.-I.}\
  \bibnamefont {Lee}},\ }\href {https://doi.org/10.1103/PhysRevLett.99.147008}
  {\bibfield  {journal} {\bibinfo  {journal} {Phys. Rev. Lett.}\ }\textbf
  {\bibinfo {volume} {99}},\ \bibinfo {pages} {147008} (\bibinfo {year}
  {2007})}\BibitemShut {NoStop}%
\bibitem [{\citenamefont {Ning}\ \emph {et~al.}(2020)\citenamefont {Ning},
  \citenamefont {Zhu}, \citenamefont {Kidd}, \citenamefont {Guan},
  \citenamefont {Wang}, \citenamefont {Mao},\ and\ \citenamefont
  {Sun}}]{Ning2020}%
  \BibitemOpen
  \bibfield  {author} {\bibinfo {author} {\bibfnamefont {J.}~\bibnamefont
  {Ning}}, \bibinfo {author} {\bibfnamefont {Y.}~\bibnamefont {Zhu}}, \bibinfo
  {author} {\bibfnamefont {J.}~\bibnamefont {Kidd}}, \bibinfo {author}
  {\bibfnamefont {Y.}~\bibnamefont {Guan}}, \bibinfo {author} {\bibfnamefont
  {Y.}~\bibnamefont {Wang}}, \bibinfo {author} {\bibfnamefont {Z.}~\bibnamefont
  {Mao}},\ and\ \bibinfo {author} {\bibfnamefont {J.}~\bibnamefont {Sun}},\
  }\href {https://doi.org/10.1038/s41524-020-00427-y} {\bibfield  {journal}
  {\bibinfo  {journal} {npj Comput. Mater.}\ }\textbf {\bibinfo {volume} {6}},\
  \bibinfo {pages} {157} (\bibinfo {year} {2020})}\BibitemShut {NoStop}%
\bibitem [{\citenamefont {Aliev}\ \emph {et~al.}(2019)\citenamefont {Aliev},
  \citenamefont {Amiraslanov}, \citenamefont {Nasonova}, \citenamefont
  {Shevelkov}, \citenamefont {Abdullayev}, \citenamefont {Jahangirli},
  \citenamefont {Orujlu}, \citenamefont {~}, \citenamefont {Mamedov},
  \citenamefont {Babanly},\ and\ \citenamefont {Chulkov}}]{Aliev2019}%
  \BibitemOpen
  \bibfield  {author} {\bibinfo {author} {\bibfnamefont {Z.~S.}\ \bibnamefont
  {Aliev}}, \bibinfo {author} {\bibfnamefont {I.~R.}\ \bibnamefont
  {Amiraslanov}}, \bibinfo {author} {\bibfnamefont {D.~I.}\ \bibnamefont
  {Nasonova}}, \bibinfo {author} {\bibfnamefont {A.~V.}\ \bibnamefont
  {Shevelkov}}, \bibinfo {author} {\bibfnamefont {N.~A.}\ \bibnamefont
  {Abdullayev}}, \bibinfo {author} {\bibfnamefont {Z.~A.}\ \bibnamefont
  {Jahangirli}}, \bibinfo {author} {\bibfnamefont {E.~N.}\ \bibnamefont
  {Orujlu}}, \bibinfo {author} {\bibfnamefont {M.~M.}\ \bibnamefont {~}},
  \bibinfo {author} {\bibfnamefont {N.~T.}\ \bibnamefont {Mamedov}}, \bibinfo
  {author} {\bibfnamefont {M.~B.}\ \bibnamefont {Babanly}},\ and\ \bibinfo
  {author} {\bibfnamefont {E.~V.}\ \bibnamefont {Chulkov}},\ }\href
  {https://doi.org/10.1016/j.jallcom.2019.03.030} {\bibfield  {journal}
  {\bibinfo  {journal} {J. Alloys Compd.}\ }\textbf {\bibinfo {volume} {789}},\
  \bibinfo {pages} {443} (\bibinfo {year} {2019})}\BibitemShut {NoStop}%
\bibitem [{\citenamefont {Liu}\ \emph {et~al.}(2020{\natexlab{b}})\citenamefont
  {Liu}, \citenamefont {Zhang}, \citenamefont {Dong}, \citenamefont {Lee},
  \citenamefont {Wei}, \citenamefont {Zhang}, \citenamefont {Chen},
  \citenamefont {Yuan}, \citenamefont {Yang},\ and\ \citenamefont
  {Qi}}]{LiuYP2020}%
  \BibitemOpen
  \bibfield  {author} {\bibinfo {author} {\bibfnamefont {Y.~P.}\ \bibnamefont
  {Liu}}, \bibinfo {author} {\bibfnamefont {Y.~J.}\ \bibnamefont {Zhang}},
  \bibinfo {author} {\bibfnamefont {J.~J.}\ \bibnamefont {Dong}}, \bibinfo
  {author} {\bibfnamefont {H.}~\bibnamefont {Lee}}, \bibinfo {author}
  {\bibfnamefont {Z.~X.}\ \bibnamefont {Wei}}, \bibinfo {author} {\bibfnamefont
  {W.~L.}\ \bibnamefont {Zhang}}, \bibinfo {author} {\bibfnamefont {C.~Y.}\
  \bibnamefont {Chen}}, \bibinfo {author} {\bibfnamefont {H.~Q.}\ \bibnamefont
  {Yuan}}, \bibinfo {author} {\bibfnamefont {Y.-f.}\ \bibnamefont {Yang}},\
  and\ \bibinfo {author} {\bibfnamefont {J.}~\bibnamefont {Qi}},\ }\href
  {https://doi.org/10.1103/PhysRevLett.124.057404} {\bibfield  {journal}
  {\bibinfo  {journal} {Phys. Rev. Lett.}\ }\textbf {\bibinfo {volume} {124}},\
  \bibinfo {pages} {057404} (\bibinfo {year} {2020}{\natexlab{b}})}\BibitemShut
  {NoStop}%
\bibitem [{\citenamefont {Balkanski}\ \emph {et~al.}(1983)\citenamefont
  {Balkanski}, \citenamefont {Wallis},\ and\ \citenamefont
  {Haro}}]{PhysRevB.28.1928}%
  \BibitemOpen
  \bibfield  {author} {\bibinfo {author} {\bibfnamefont {M.}~\bibnamefont
  {Balkanski}}, \bibinfo {author} {\bibfnamefont {R.~F.}\ \bibnamefont
  {Wallis}},\ and\ \bibinfo {author} {\bibfnamefont {E.}~\bibnamefont {Haro}},\
  }\href {https://doi.org/10.1103/PhysRevB.28.1928} {\bibfield  {journal}
  {\bibinfo  {journal} {Phys. Rev. B}\ }\textbf {\bibinfo {volume} {28}},\
  \bibinfo {pages} {1928} (\bibinfo {year} {1983})}\BibitemShut {NoStop}%
\bibitem [{\citenamefont {Men\'endez}\ and\ \citenamefont
  {Cardona}(1984)}]{PhysRevB.29.2051}%
  \BibitemOpen
  \bibfield  {author} {\bibinfo {author} {\bibfnamefont {J.}~\bibnamefont
  {Men\'endez}}\ and\ \bibinfo {author} {\bibfnamefont {M.}~\bibnamefont
  {Cardona}},\ }\href {https://doi.org/10.1103/PhysRevB.29.2051} {\bibfield
  {journal} {\bibinfo  {journal} {Phys. Rev. B}\ }\textbf {\bibinfo {volume}
  {29}},\ \bibinfo {pages} {2051} (\bibinfo {year} {1984})}\BibitemShut
  {NoStop}%
\bibitem [{\citenamefont {Yang}\ \emph {et~al.}(2021)\citenamefont {Yang},
  \citenamefont {Xu}, \citenamefont {Zhu}, \citenamefont {Niu}, \citenamefont
  {Xu}, \citenamefont {Peng}, \citenamefont {Cheng}, \citenamefont {Jia},
  \citenamefont {Huang}, \citenamefont {Xu}, \citenamefont {Lu},\ and\
  \citenamefont {Ye}}]{PhysRevX.11.011003}%
  \BibitemOpen
  \bibfield  {author} {\bibinfo {author} {\bibfnamefont {S.}~\bibnamefont
  {Yang}}, \bibinfo {author} {\bibfnamefont {X.}~\bibnamefont {Xu}}, \bibinfo
  {author} {\bibfnamefont {Y.}~\bibnamefont {Zhu}}, \bibinfo {author}
  {\bibfnamefont {R.}~\bibnamefont {Niu}}, \bibinfo {author} {\bibfnamefont
  {C.}~\bibnamefont {Xu}}, \bibinfo {author} {\bibfnamefont {Y.}~\bibnamefont
  {Peng}}, \bibinfo {author} {\bibfnamefont {X.}~\bibnamefont {Cheng}},
  \bibinfo {author} {\bibfnamefont {X.}~\bibnamefont {Jia}}, \bibinfo {author}
  {\bibfnamefont {Y.}~\bibnamefont {Huang}}, \bibinfo {author} {\bibfnamefont
  {X.}~\bibnamefont {Xu}}, \bibinfo {author} {\bibfnamefont {J.}~\bibnamefont
  {Lu}},\ and\ \bibinfo {author} {\bibfnamefont {Y.}~\bibnamefont {Ye}},\
  }\href {https://doi.org/10.1103/PhysRevX.11.011003} {\bibfield  {journal}
  {\bibinfo  {journal} {Phys. Rev. X}\ }\textbf {\bibinfo {volume} {11}},\
  \bibinfo {pages} {011003} (\bibinfo {year} {2021})}\BibitemShut {NoStop}%
\bibitem [{\citenamefont {He}(2020)}]{npjQuantumMaterials5902020}%
  \BibitemOpen
  \bibfield  {author} {\bibinfo {author} {\bibfnamefont {K.}~\bibnamefont
  {He}},\ }\href {https://doi.org/10.1038/s41535-020-00291-5} {\bibfield
  {journal} {\bibinfo  {journal} {npj Quantum Mater.}\ }\textbf {\bibinfo
  {volume} {5}},\ \bibinfo {pages} {90} (\bibinfo {year} {2020})}\BibitemShut
  {NoStop}%
\end{thebibliography}
%

\end{document}